\documentclass{cplslarge}%
\usepackage[authoryear]{natbib}
\usepackage{makeidx}

\usepackage[switch]{lineno}

\usepackage[bookmarks = true, bookmarksnumbered = true, pdfpagemode =None, pdfstartview = FitH, pdfpagelayout = SinglePage, colorlinks = true, urlcolor = red, citecolor = blue]{hyperref}

\usepackage{graphics}

\makeindex

\global\chapterreferencefalse
\usepackage[sectionbib]{chapterbib}

\begin{document}
\frontmatter
\mainmatter
  
\author[Fletcher, Greathouse, Guerlet, Moses \&\ West]{l.n. fletcher$^1$, t.k. greathouse$^2$, s. guerlet$^3$, j.i. moses$^4$ and r.a. west$^5$}
\chapter{Saturn's seasonally changing atmosphere:  thermal structure, composition and aerosols}

\footnotesize
$^1$ Department of Physics and Astronomy, University of Leicester, University Road, Leicester, LE1 7RH, UK.\\
$^2$ Southwest Research Institute, Department of Space Science, Space Science and Engineering Division, 6220 Culebra Road, San Antonio, TX 78238-5166, USA.\\
$^3$ Laboratoire de Meteorologie Dynamique / CNRS / Univ. Paris 6, 4 Place Jussieu, 75252 Paris, France.\\
$^4$ Space Sciences Institute, 4750 Walnut St, Suite 205, Boulder, CO 80301, USA.\\
$^5$ NASA Jet Propulsion Laboratory, 4800 Oak Grove Drive, Pasadena, CA 91109, USA.

\section*{Copyright Notice}
The Chapter, Saturn's Seasonally Changing Atmosphere: Thermal Structure, Composition and Aerosols, is to be published by Cambridge University Press as part of a multi-volume work edited by Kevin Baines, Michael Flasar, Norbert Krupp, and Thomas Stallard, entitled ``Saturn in the 21st Century" (`the Volume')
 
\copyright  ~in the Chapter, L.N. Fletcher, T.K. Greathouse, S. Guerlet, J.I. Moses and R.A. West

\copyright ~in the Volume, Cambridge University Press
 
NB: The copy of the Chapter, as displayed on this website, is a draft, pre-publication copy only. The final, published version of the Chapter will be available to purchase through Cambridge University Press and other standard distribution channels as part of the wider, edited Volume, once published. This draft copy is made available for personal use only and must not be sold or re-distributed.

\normalsize

\section*{Abstract}

The longevity of Cassini's exploration of Saturn's atmosphere (a third of a Saturnian year) means that we have been able to track the seasonal evolution of atmospheric temperatures, chemistry and cloud opacity over almost every season, from solstice to solstice and from perihelion to aphelion.  Cassini has built upon the decades-long ground-based record to observe seasonal shifts in atmospheric temperature, finding a thermal response that lags behind the seasonal insolation with a lag time that increases with depth into the atmosphere, in agreement with radiative climate models.  Seasonal hemispheric contrasts are perturbed at smaller scales by atmospheric circulation, such as belt/zone dynamics, the equatorial oscillations and the polar vortices. Temperature asymmetries are largest in the middle stratosphere and become insignificant near the radiative-convective boundary.  Cassini has also measured southern-summertime asymmetries in atmospheric composition, including ammonia (the key species for the topmost clouds), phosphine and para-hydrogen (both disequilibrium species) in the upper troposphere; and hydrocarbons deriving from the UV photolysis of methane in the stratosphere (principally ethane and acetylene).   These chemical asymmetries are now altering in subtle ways due to (i) the changing chemical efficiencies with temperature and insolation; and (ii) vertical motions associated with large-scale overturning in response to the seasonal temperature contrasts.  Similarly, hemispheric contrasts in tropospheric aerosol opacity and coloration that were identified during the earliest phases of Cassini's exploration have now reversed, suggesting an intricate link between the clouds and the temperatures.  Finally, comparisons of observations between Voyager and Cassini (both observing in early northern spring, one Saturn year apart) show tantalising suggestions of non-seasonal variability.  Disentangling the competing effects of radiative balance, chemistry and dynamics in shaping the seasonal evolution of Saturn's temperatures, clouds and composition remains the key challenge for the next generation of observations and numerical simulations.

\section{Introduction}
\label{intro}

\begin{figure*}%
\begin{center}
\figurebox{7in}{}{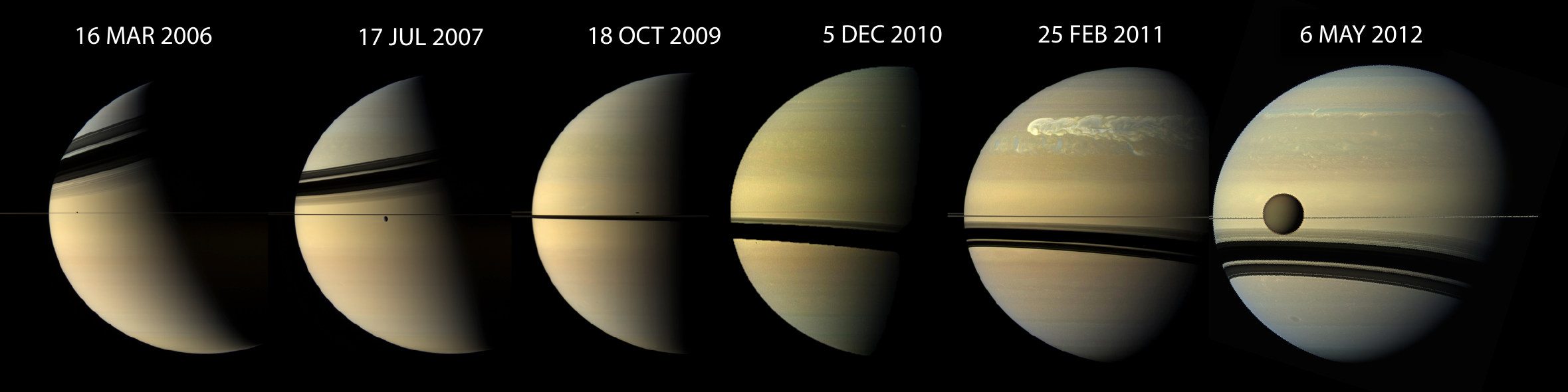}
\caption{Saturn's changing insolation from 2006 to 2012, three years on either side of the northern spring equinox. The colors of Saturn's tropospheric clouds and hazes can be seen shifting as northern winter becomes northern spring.  Compiled from Cassini images courtesy of NASA/JPL-Caltech.}
\label{saturn_montage}
\end{center}
\end{figure*}

We can achieve a greater understanding of any complex system by studying how that system evolves with time.  Saturn, with its $26.7^\circ$ ÒEarth-likeÓ axial tilt, 29.5-Earth-year orbital period and orbital eccentricity of 0.057 (perihelion near northern winter solstice, aphelion near northern summer solstice), is our closest and best example of a seasonally-variable giant planet atmosphere, in contrast with Jupiter (negligible seasonal influences from the $3.1^\circ$ tilt), Uranus (extreme seasonal contrasts from the $98^\circ$ tilt) and Neptune (slow evolution due to the 165-year period).  Furthermore, Cassini has provided our best opportunity in a generation to study the seasonal evolution of a giant planet atmosphere and the influence of temporal variations in sunlight on the atmospheric temperatures, clouds and chemistry.  Saturn completed 13.5 orbits of the Sun between Galileo's first glimpses of the ringed planet and its ``strange appendages'' (July 1610) and the Cassini spacecraft's arrival at Saturn orbit in June 2004.  Studies of seasonally-changing atmospheric properties have only been possible for the past four decades (1.3 Saturn years, coinciding with the improved capabilities of ground-based infrared remote sensing), with the Cassini orbiter providing reconnaissance for the last third of a Saturn year (Fig. \ref{saturn_montage}).  In this chapter we review our current knowledge of Saturn's temporally-variable thermal structure, composition and aerosols, from the churning tropospheric cloud decks to the middle atmosphere.  

What properties might we expect to be time-variable on a giant planet?  Atmospheric temperatures are governed by a delicate balance between Saturn's internal heat source and heating from the Sun.  Both diabatic (balance between radiative heating and cooling) and adiabatic (atmospheric motions redistributing energy vertically and horizontally) forcings govern the spatial variations of temperature, so that we might expect warm summers and cool winters albeit with a phase lag compared to the solstices due to the atmospheric inertia.  The seasonal diabatic heating will generate hemispheric temperature gradients that are superimposed on the belt/zone structure, which then drive atmospheric transport to redistribute excess energy.  These hemispheric temperature contrasts should remain in balance with Saturn's zonal wind system (via the thermal wind relation), such that vertical shears on the zonal jets could vary with time.  Saturn's chemical composition may also vary with season, as variations in ultraviolet insolation drive ionization and photodissociation rates, which in turn govern the populations of hydrocarbons and hazes derived from methane photolysis in the stratosphere, and the distribution of key volatiles (e.g., NH$_3$) and disequilibrium species (e.g., PH$_3$) in the upper troposphere.  Photochemically-produced hazes could sediment downward to serve as cloud-condensation nuclei for condensible volatiles, which in turn could cause aerosol and cloud properties to vary with time (e.g., Fig. \ref{saturn_montage}).  Finally, dynamic phenomena in the weather layer (vortices, storms, plumes and waves) respond to modifications of atmospheric stratification, so that seasons could help modulate meteorological activity.  

Cassini's longevity, coupled with a long baseline of ground-based observing, allows us to monitor each of these processes during a Saturn year to provide a four-dimensional understanding of Saturn's troposphere and stratosphere.   Seasons are indicated by the  planetocentric solar longitude ($L_s$), from $0^\circ$ at northern spring equinox (1980, 2009), to $90^\circ$ at the northern summer solstice (1987, 2017), $180^\circ$ at northern autumnal equinox (1995) and $270^\circ$ at the northern winter solstice (2002).  Saturn's southern summers receive greater insolation than northern summers (perihelion occurs near $L_s=280^\circ$, July 2003; aphelion at $L_s=100^\circ$, April 2018).  In addition to ground-based remote sensing since the mid-1970s (southern summer), Saturn's seasonal asymmetries have been observed by four visiting spacecraft:  observations by Pioneer 11 and Voyagers 1 and 2 were clustered around northern spring equinox (1979-1981); Cassini entered Saturn orbit in June 2004, two years after southern summer solstice ($L_s=293^\circ$) and aims to complete its mission at northern summer solstice ($L_s=93^\circ$ in September 2017).  

\adjustfigure{40pt}

This chapter is organized as follows:   Section \ref{temp} reviews investigations of Saturn's tropospheric and stratospheric temperature field, comparing them to climate models to understand the influence and variability of atmospheric circulation.  In Section \ref{chem} we review observations and chemical modeling of the spatial distributions and variability of key atmospheric species.  Section \ref{clouds} reviews the characteristics of Saturn's clouds and hazes, focusing on their time-variable properties, before we review unanswered questions in Section \ref{conclude}.  Planetographic latitudes are assumed unless otherwise stated.  We confine our discussion to Saturn's seasonally variable troposphere and stratosphere; the thermosphere and ionosphere will be discussed in Chapter 9.

\section{Seasonally-evolving thermal structure}
\label{temp}

\subsection{Pre-Cassini studies}

Saturn's temperature structure in the cloud-forming region and the lower troposphere is expected to follow an adiabatic gradient, with the lapse rate dominated by the heat capacity of the hydrogen-helium atmosphere but with small contributions from latent heat released by the condensation of volatile species (NH$_3$, NH$_4$SH and H$_2$O) and lagged conversion between the two different spin isomers (ortho- and para-H$_2$) of molecular hydrogen \citep[see the review by][and Section \ref{sec_parah2}]{84ingersoll}.  At lower pressures above the radiative-convective boundary \citep[350-500 mbar,][]{07fletcher_temp}, the atmospheric opacity drops sufficiently to allow efficient cooling by radiation and the temperatures deviate from the adiabat, becoming more stably stratified in the upper troposphere towards the temperature minimum (the tropopause near 80 mbar).  Atmospheric heating due to short-wavelength sunlight absorption by methane and aerosols causes temperatures to rise again in the stratosphere, balanced by long-wavelength cooling from methane (upper stratosphere), ethane and acetylene (middle and lower stratosphere), and the collision-induced hydrogen-helium continuum (upper troposphere and lower stratosphere).  It is the seasonal dependence of the atmospheric heating (and cooling, via chemical changes) that causes Saturn's upper tropospheric and stratospheric temperatures to vary considerably with season.

Seasonal temperature variations were observed as asymmetries in emission measured at thermal infrared wavelengths, first detected in ground-based images from stratospheric ethane near 12 $\mu$m \citep{75gillett, 75rieke} and methane near 8 $\mu$m \citep{78tokunaga} during southern summertime conditions, revealing enhanced emission from the southern pole.  Tropospheric contrasts in the 17-23 $\mu$m region were still present but more muted \citep{78caldwell, 78tokunaga}, consistent with a seasonal response that becomes weaker with depth.   The first spacecraft measurements of Saturn's temperatures by Pioneer 11 (1979, $L_s=354^\circ$) revealed no tropospheric thermal asymmetries between 10$^\circ$N and 30$^\circ$S just before the northern spring equinox \citep{80orton}.  However, inversions of Voyager 1 (1980, $L_s=8.6^\circ$) and 2 (1981, $L_s=18.2^\circ$) IRIS 14-50 $\mu$m spectra revealed tropospheric temperature asymmetries \citep{81hanel, 82hanel, 83conrath, 98conrath}, with the north cooler than the south at 150-200 mbar shortly after northern spring equinox.  The equinoctial timing of these observations suggested a delay between increased insolation (the solar forcing) and the atmospheric response \citep{79cess} as a consequence of the high thermal inertia of the atmosphere.  At tropospheric depths of 500-700 mbar this inertia, and hence the lagged response to the solar forcing, becomes so large that Saturn's northern and southern hemisphere temperatures do not show significant asymmetries \citep{83conrath}.  Seasonal amplitudes are largest in the stratosphere, as we shall describe below.

\begin{figure}%
\begin{center}
\figurebox{3.5in}{}{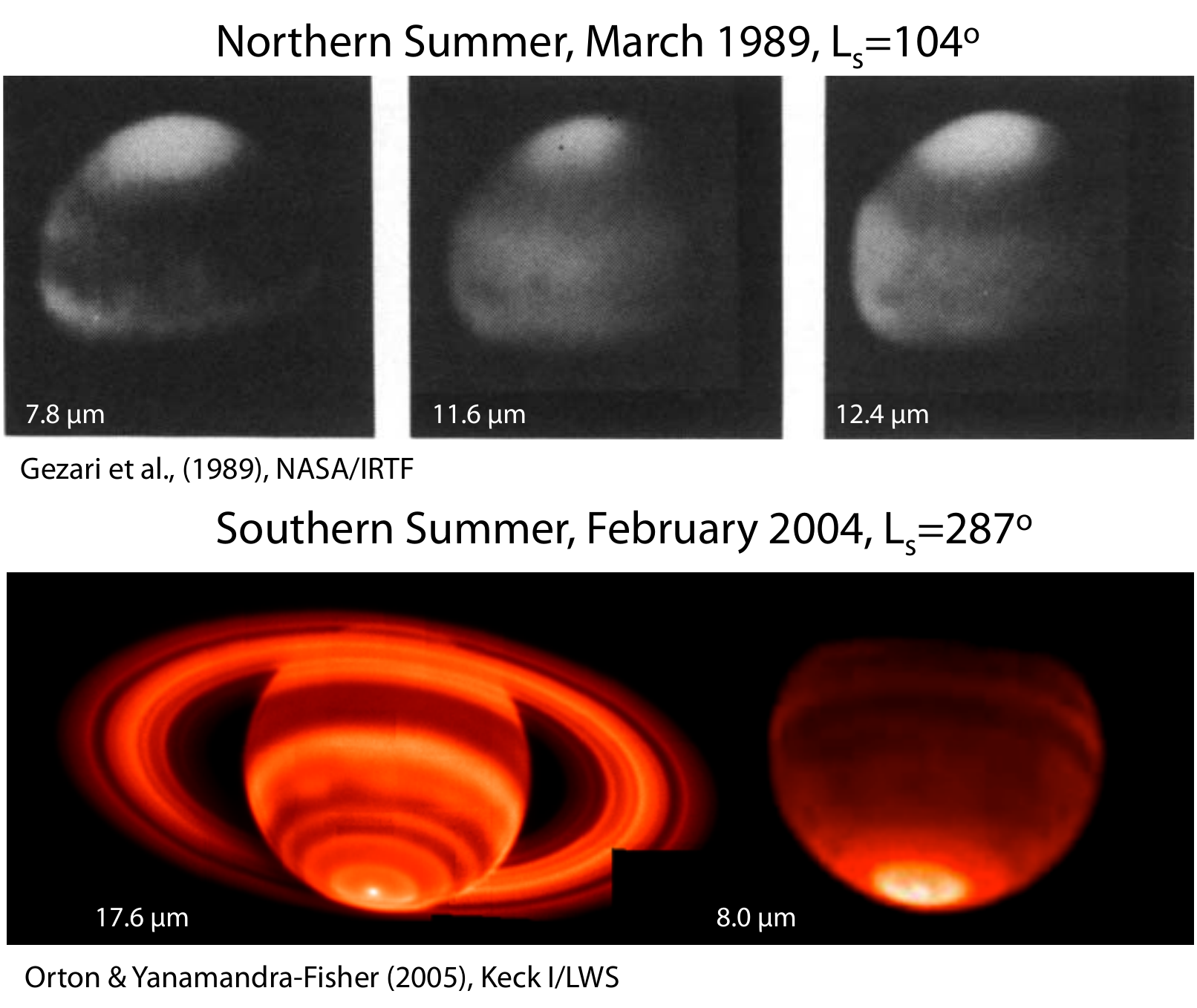}
\caption{Stratospheric thermal emission imaged by ground-based facilities in March 1989 \citep[near northern summer solstice,][]{89gezari} and February 2004 \citep[near southern summer solstice,][]{05orton}, showing enhanced emission from the summer pole and similarities in the seasonal response.}
\label{thermalimages}
\end{center}
\end{figure}

Twenty-three years would pass before another spacecraft reached the Saturn system, but ground-based studies continued to provide insights into Saturn's seasons, particularly with the advent of 2D mid-infrared detector technologies and high-resolution spectroscopy \citep[see the review by][]{09orton}.  Saturn's north polar regions exhibited enhanced methane and ethane emission by early northern summer \citep[observations by][in March 1989, $L_s=104^\circ$, Fig. \ref{thermalimages}]{89gezari} just like the south pole had in southern summer, although this enhanced emission was not readily observable in the troposphere \citep[observations in December 1992, $L_s=146^\circ$ by][]{00ollivier}.  When southern summer returned in the early 2000s, \citet{05greathouse} used high-resolution CH$_4$ emission spectroscopy to derive a stratospheric temperature asymmetry at southern summer solstice (2002, $L_s=268^\circ$, the south pole 10 K warmer than the equator) and an asymmetry that weakened with increasing depth; while \citet{05orton} presented high resolution 7-25 $\mu$m images that revealed the southern summer hemisphere in exquisite detail (February 2004, $L_s=287^\circ$, Fig. \ref{thermalimages}).  Specifically, they observed a 15 K temperature contrast from the equator to the south pole at 3 mbar, a sharp temperature gradient near $70^\circ$S (the edge of the south polar warm hood) and an intense tropospheric hotspot associated with the south polar cyclone poleward of $87^\circ$S.  These observations confirmed that Cassini would observe a stark asymmetry between the northern winter and southern summer hemispheres upon arrival in 2004.  

\subsection{Cassini observations in southern summer}
Cassini's great advantage is the ability to view both the northern and southern hemispheres near-simultaneously, whereas Saturnian winter is forever hidden to an Earth-based observer.  Tropospheric and stratospheric temperatures are provided via a combination of infrared remote sensing, radio occultation and ultraviolet stellar occultations.  Nadir 7-1000 $\mu$m spectroscopy with the Composite Infrared Spectrometer \citep[CIRS,][]{04flasar} measures tropospheric temperatures from the tropopause down to the topmost cloud decks (approximately 80-800 mbar) and stratospheric temperatures (from methane emission) in the 0.5-5.0 mbar range \citep{05flasar, 07fletcher_temp}.   CIRS limb observations \citep{08fouchet, 09guerlet} complement the vertical coverage of nadir observations by constraining the stratospheric temperature profile between 20 mbar and 1 $\mu$bar with a vertical resolution of 1-2 scale heights, at the expense of poorer spatial and temporal coverage. Radio-occultations of Cassini by Saturn \citep{11schinder}, available at a limited number of latitudes, achieve the highest vertical resolution (6-7 km, a tenth of a scale height) and constrain the temperature-pressure profile between 1500 mbar and 0.1 mbar. 

\begin{figure}%
\begin{center}
\figurebox{3.3in}{}{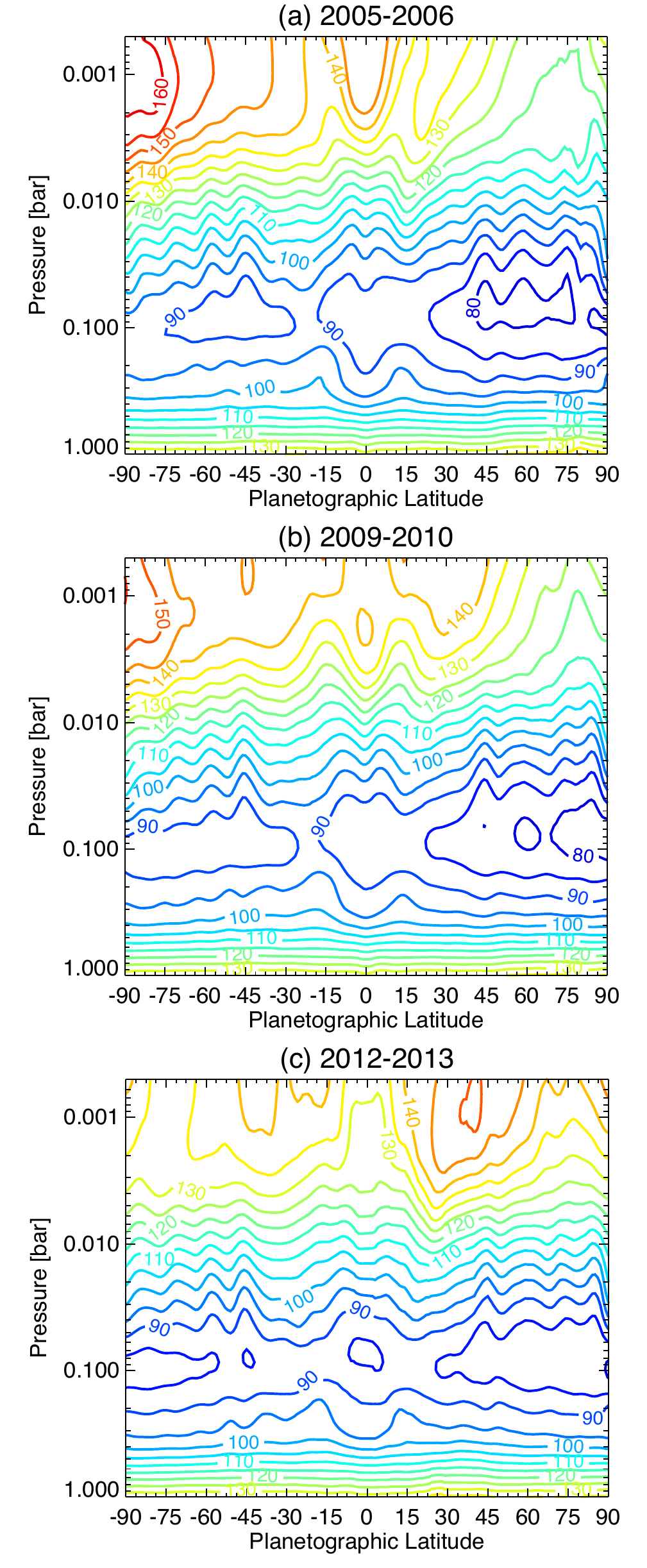}
\caption{Zonal mean temperatures derived from Cassini/CIRS nadir infrared spectroscopy at three different epochs, representing a snapshot of the temperature field in (a) late southern summer (2005-2006, $L_s\approx313^\circ$); (b) northern spring equinox (2009-2010, $L_s\approx5^\circ$) and (c) early northern spring (2013-2014, $L_s\approx52^\circ$).  Adapted and updated from \citet{10fletcher_seasons} and \citet{15fletcher_poles}.}
\label{temp_profile}
\end{center}
\end{figure}

Cassini's prime mission provided a snapshot of Saturn's hemispheric temperature asymmetries in late southern summer \citep[2004-2008, $L_s=293-345^\circ$,][]{05flasar, 07fletcher_temp, 08fletcher_poles, 09guerlet}, as shown in Fig. \ref{temp_profile}(a). The summer pole was found to be 40 K warmer than the winter pole at 1 mbar, with the contrast decreasing with increasing pressure \citep{07fletcher_temp}.  Intriguingly, the latitudinal asymmetry appeared to be smaller ($\approx24$ K) at 0.1 mbar and smaller still at 0.01 mbar\citep{09guerlet}.  The tropopause was around 10 K cooler (and slightly higher in altitude) in the winter hemisphere than the summer hemisphere \citep{07fletcher_temp}.  Below the tropopause, the lapse rate increased with depth until reaching the dry adiabat at approximately 350-500 mbar \citep{85lindal, 07fletcher_temp}, likely indicating the location of the radiative-convective boundary that separates the stably stratified upper troposphere from the convective deeper troposphere.  This lapse rate change occurred at higher pressures (400-500 mbar) in the summer hemisphere than in the northern hemisphere (350-450 mbar), due to the greater penetration of solar heating in the southern hemisphere.  Temperature asymmetries became negligible for $p>500$ mbar.  

Between the radiative-convective boundary and the tropopause, inversions of far-infrared spectra revealed an inflection point in the tropospheric temperature structure in the 100-300 mbar region.  This perturbation was referred to as the `knee' \citep{07fletcher_temp}, suggestive of heating that was localised in altitude within Saturn's tropospheric haze.  It was first noted in Voyager/IRIS retrievals at equatorial and southern latitudes \citep{81hanel} and in Voyager radio occultations at $3^\circ$S and $74^\circ$S \citep{85lindal}.  The pressure level and magnitude of this temperature perturbation varied strongly with latitude, being enhanced in southern summer and weak or absent in the northern winter hemisphere.  The knee was higher and weaker over the equator, and showed local maxima at 15$^\circ$N and 15$^\circ$S associated with the warm equatorial belts (see Fig. \ref{cloud_compare} at the end of this chapter). \citet{07fletcher_temp} concluded that this was a radiative effect due to solar absorption by aerosols in the upper troposphere, and explained the asymmetry in terms of both seasonal insolation and the variable distributions of tropospheric aerosols \citep[later confirmed by radiative climate modeling by][]{12friedson, 14guerlet}.  The relationship between the `knee' and the upper tropospheric haze is discussed in Section \ref{clouds}.  


Saturn's temperature distribution in Fig. \ref{temp_profile} reveals the influence of dynamics as well as radiative balance.  The global temperature asymmetries are superimposed onto small-scale latitudinal contrasts between the cool zones (anticyclonic shear regions equatorward of prograde jets) and warmer belts (cyclonic shear regions poleward of prograde jets) \citep{83conrath, 07fletcher_temp}.  Note that the correlations between this belt/zone structure (defined in terms of the zonal jets and tropospheric temperatures) and the cloud reflectivity is not as clear-cut as for Jupiter.  For example, bands of low reflectivity are often narrow and located close to the prograde jet peaks \citep[e.g.,][]{06vasavada}, and do not appear correlated with the thermal structure.  Saturn's polar troposphere features long-lived cyclonic `hot spots' located directly at each pole irrespective of season, and the northern hexagonal jet at 77$^\circ$N is related to a hexagonal warm polar belt poleward of this latitude \citep{08fletcher_poles}. At lower pressures, the polar stratosphere features a warm `polar hood' in summer that is absent in winter, one of the most extreme examples of a seasonal phenomenon on Saturn \citep{08fletcher_poles}.  In Saturn's tropical stratosphere, contrasts between the equator and neighboring latitudes are observed to oscillate with time and altitude. This `semi-annual oscillation' in the thermal structure implies a strong vertical shear of the zonal wind at the equator and is reminiscent of the Quasi-Biennal Oscillation in the Earth's stratosphere, a dynamical phenomenon driven by wave-zonal flow interactions \citep{08fouchet, 08orton_qxo}.

\subsection{Seasonal evolution of temperatures}

Saturn's thermal structure during southern summer was reviewed by \citet{09delgenio}, but Cassini has since revealed how the global temperature structure has evolved with season through northern spring equinox.  \citet{10fletcher_seasons} used CIRS observations from 2004 to 2009 ($L_s=297-358^\circ$) to determine Saturn's upper tropospheric and stratospheric temperature variability, finding (i) stratospheric warming of northern mid-latitudes by 6-10 K at 1 mbar as they emerged from ring shadow into springtime conditions; (ii) southern cooling by 4-6 K (both at mid-latitudes and within the south polar stratospheric hood poleward of $70^\circ$S) and a resulting decrease in the 40-K asymmetry between the hemispheres that was present in 2004; and (iii) a tropospheric response to the seasonal insolation shifts that seemed to be larger at the locations of the broadest retrograde jets.  The `flattening' of the summertime temperature asymmetry (by Saturn's equinox, northern and southern mid-latitude 1-mbar temperatures were both in the 140-145 K range) followed the expectations of a radiative climate model \citep[][and see below]{10greathouse}, albeit perturbed by the equatorial oscillation, polar vortex dynamics and the belt/zone structure.  \citet{13sinclair} extended this analysis to include observations in 2010 ($L_s=15^\circ$) and observed the continued northern warming and southern cooling, particularly intense within Saturn's south polar region ($\approx17$ K between 2005 and 2010).  At higher stratospheric altitudes, \citet{15sylvestre} extended the analysis of \citet{09guerlet} by considering limb spectroscopy in 2010-2012, finding a seasonal trend consistent with the nadir data at 1 mbar, but with smaller variations at lower pressures.  For example, mid-latitude temperatures near 0.1 mbar have remained approximately constant between 2005 and 2010.  

\begin{figure}%
\begin{center}
\figurebox{3.2in}{}{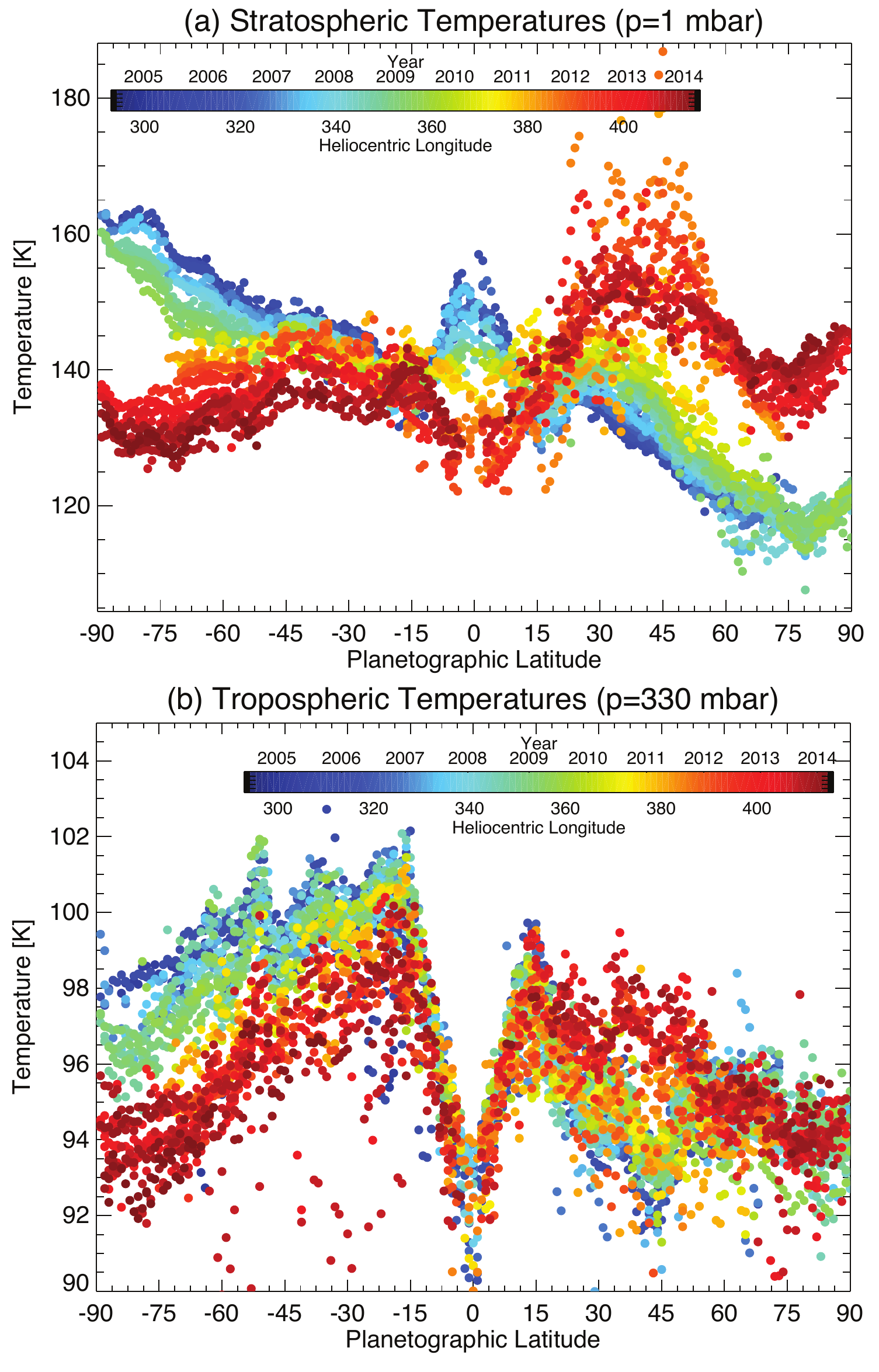}
\caption{Zonal mean temperatures in the stratosphere (1 mbar) and troposphere (330 mbar) as derived from nadir Cassini/CIRS spectra over the duration of the Cassini mission.  Note that the absence of high-latitude measurements between 2010 and 2012 is due to Cassini's near-equatorial orbit at that time, preventing nadir observations of the poles.  Updated from \citet{10fletcher_seasons} and \citet{15fletcher_poles}.}
\label{seasonal_temp}
\end{center}
\end{figure}

\begin{figure*}%
\begin{center}
\figurebox{7.in}{}{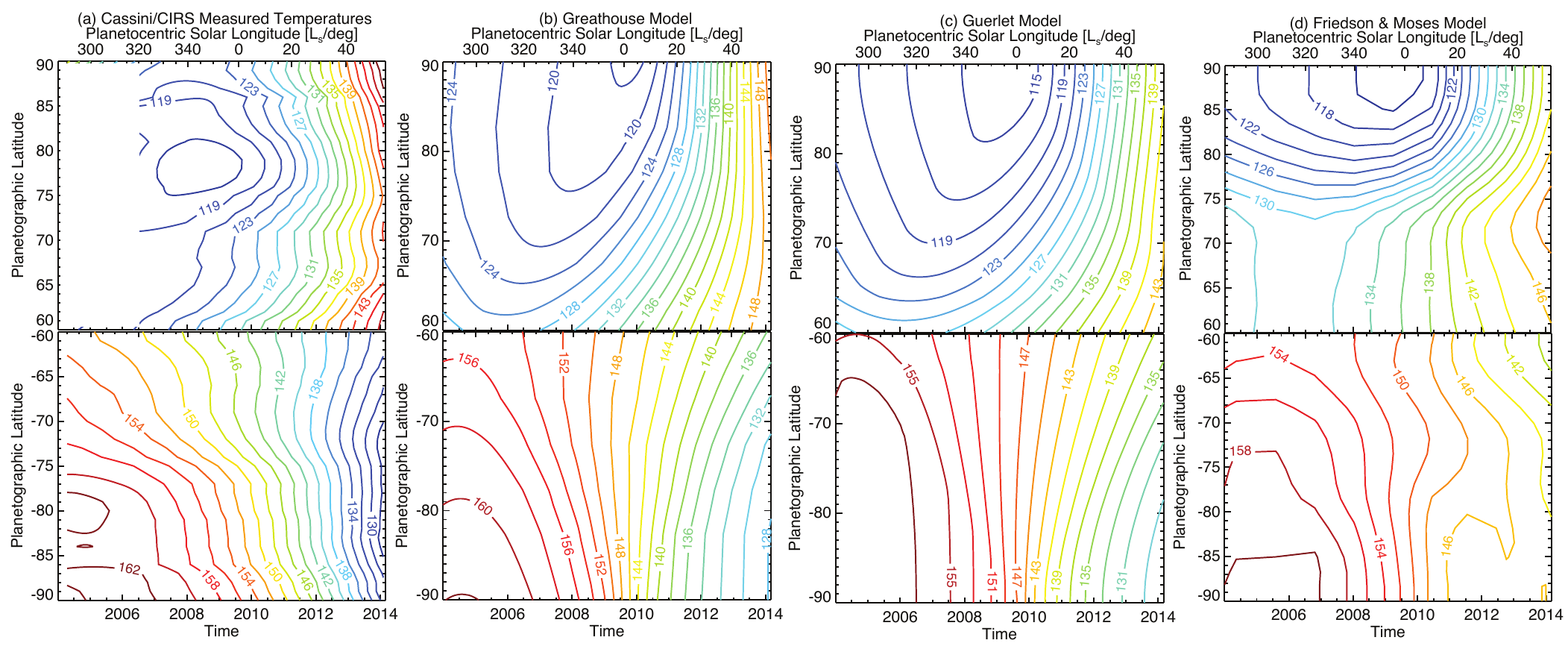}
\caption{Temperatures at 1 mbar derived in the polar regions of Saturn \citep{15fletcher_poles}, exposed to the most severe insolation changes over a saturnian year.  Cassini measurements are compared to the model predictions of \citet{10greathouse},  \citet{14guerlet} and \citet{12friedson}.  Although discrepancies in the absolute temperatures are evident, the timing of the polar maxima/minima in temperatures is reasonably well reproduced.}
\label{polartemp}
\end{center}
\end{figure*}

Cassini's monitoring of the seasonal temperatures has now extended well into northern spring, and most recently \citet{15fletcher_poles} extended the work of \citet{10fletcher_seasons} and \citet{13sinclair} to 2014 ($L_s=56^\circ$).  The zonal mean temperature as a function of latitude and pressure is shown for three epochs in Fig. \ref{temp_profile}, and all CIRS nadir retrievals are shown as a function of time in Fig. \ref{seasonal_temp}.   Focusing on the highest latitudes, tropospheric contrasts between the cool polar zones ($80-85^\circ$ latitude) and warm polar belts (near $75-80^\circ$ latitude) have varied over the ten-year span of observations, indicating changes to the vertical shear on the zonal jets via the thermal wind equation (Fig. \ref{temp_profile}).  The warm south polar stratosphere has cooled dramatically by $\approx5$ K/yr, mirrored by warming of a similar magnitude in the north.  However, while the south polar region was isolated by a strong thermal gradient near $75^\circ$S during the height of summer, no similar boundary was apparent near $75^\circ$N during spring despite rising temperatures towards the north pole at $L_s=56^\circ$ (the last published data), suggesting that the northern summer vortex has yet to form \citep[see Fig. \ref{polartemp}, Chapter 12 and][]{15fletcher_poles}.  The peak stratospheric warming in the north was occurring at lower pressures (0.5-1.0 mbar) than the peak stratospheric cooling in the south (1-3 mbar).  Fig. \ref{polartemp} demonstrates that north polar minima in stratospheric temperatures were detected in 2008-2010 (lagging one season, or 6-8 years, behind winter solstice); south polar maxima appear to have occurred before the start of the Cassini observations (1-2 years after summer solstice).

\subsection{Seasonal climate modeling}

Seasonal climate modeling is an attempt to create models that accurately predict the temporal evolution of a planet's thermal structure as a function of altitude, latitude and longitude.  Diurnal temperature variations are not expected due to Saturn's high thermal inertia, the relatively low amount of solar forcing at Saturn's distance ($\approx100$ times less than at Earth), and Saturn's fast rotation (10 hour 39 minute long days), and current models bear this out \citep{08greathouse_agu,14guerlet}.  This fact allows modelers to reduce the complexity of the problem and focus on the temporal evolution of temperatures as a function of altitude and latitude only, calling then for detailed 2-D time variable models.  

The cooling of Saturn's stratosphere is dominated by the radiative emissions from C$_2$H$_2$ and C$_2$H$_6$ at pressures lower than 5 mbar, along with some cooling due to the $\nu_4$ band of CH$_4$ near 8 $\mu$m.   Cooling due to other hydrocarbons is estimated to account for no more than 5\% of the total radiative cooling rate \citep{14guerlet}.  Tropospheric and lower stratospheric cooling occurs via emission from the H$_2$-H$_2$ and H$_2$-He collision-induced continuum.  Atmospheric heating is due to the absorption of sunlight primarily by CH$_4$ and aerosols.  C$_2$H$_2$ and C$_2$H$_6$ are both byproducts of CH$_4$ photochemistry (for details see Section \ref{chem}) which is initiated at the top of the atmosphere just below the CH$_4$ homopause.  Since C$_2$H$_2$ and C$_2$H$_6$ are the dominant radiative coolants, and their distribution controls the extent and direction of radiant energy to space, their distribution can alter stratospheric temperatures, which in turn can induce circulation patterns that serve to redistribute the molecules. While this complicated interchange continues, ongoing photochemical processes are constantly making their own adjustments to the abundances of C$_2$H$_2$ and C$_2$H$_6$.  This interconnectedness means that to produce a truly accurate seasonal climate model one needs to include accurate calculations of the absorption and emission of radiation, photochemistry, and dynamics.  Any one of these would make for a complex model on its own and the combination of all three is a worthy goal.  While this level of complexity is currently not achieved, we show below that much has been accomplished in all three disciplines, and the most recent incarnations of seasonal models are moving ever closer to this goal.
\adjustfigure{60pt}

The first generation of seasonal climate models for Saturn were inspired by the first observations of seasonal temperature asymmetries in the 1970s and Pioneer-Voyager epoch. Radiative-convective equilibrium models in the 1970s \citep{77caldwell, 77tokunaga, 84appleby} showed that the solar absorption by methane in the visible and near-infrared could explain the temperature inversion identified from ethane limb brightening \citep{75gillett}.  Early models of Saturn's stratospheric temperature response \citep{79cess, 80carlson} were extended into the troposphere by \citet{84bezard}, and indicated that an optimal fit to the measured tropospheric temperatures required an additional source of opacity (potentially from tropospheric aerosols). \citet{85bezard} incorporated a non-gray treatment of radiative transfer to construct a more sophisticated model, accounting for seasonal changes in solar forcing, ring obscuration and planetary oblateness.  Though they were limited by both the lack of some lab measurements \citep[several near-infrared bands of methane had yet to be measured in the lab,][]{85bezard} and data on Saturn (detailed measurements of temperatures versus altitude, latitude, and time along with the variations of mixing ratio for C$_2$H$_2$ and C$_2$H$_6$ as a function of time, altitude and latitude), they were able to produce models that to first order reproduced early ground based observations and much of the data retrieved by Voyager 1 and 2, including the observation that the thermal inertia increased with depth such that there was no seasonal modulation of the deepest atmospheric temperatures. These models were one-dimensional in nature, requiring the user to model each latitude of interest and then compile the results to show how the global temperatures changed over time.  \citet{90conrath} produced more sophisticated radiative-convective-dynamical models of Saturn's troposphere and stratosphere for a comparison to Voyager results, and \citet{92barnet} included the effects of ring shadowing and ring thermal emission.  

The radiative climate models have evolved with time as (i) the spectroscopic parameters of opacity sources (methane, hydrocarbons and hazes) have become better constrained; and (ii) the spatial distributions of the key infrared coolants (ethane and acetylene) have become better known; and (iii) the vertical distribution of clouds and hazes, which have a substantial contribution to the radiative budget, have been revealed.  The richness and complexity in the Cassini seasonal dataset prompted the evolution of a new generation of radiative climate models, taking advantage of improvements in laboratory measurements and computational capabilities.  The first such model \citep[][hereafter TG]{10greathouse} was a purely radiative seasonal model of Saturn's stratosphere, with the capability of assuming any vertical, meridional, or temporal variation of the key hydrocarbon coolants.  This model is one dimensional in nature, one latitude over time, requiring multiple runs to compile results from different latitudes into a 2-D representation of Saturn's stratospheric seasonal evolution.  However, as initially planned, this model was absorbed as a module within the Explicit Planetary Isentropic-Coordinate Global Circulation Model (EPIC GCM), allowing EPIC to accurately calculate the radiative heating and cooling rates while accounting for dynamics \citep{10dowling}.  The second model, the Outer Planet General Circulation Model (OPGCM), is a stratospheric and tropospheric seasonal dynamical model which efficiently and accurately accounts for the radiative heating and cooling while also tracking circulation caused by the seasonal forcing in 3-dimensions \citep[][hereafter FM]{12friedson}.  Finally, the most recent seasonal model is the stratospheric and tropospheric radiative-convective model produced by \citet{14guerlet} (hereafter SG).  

\begin{figure}%
\begin{center}
\figurebox{3.2in}{}{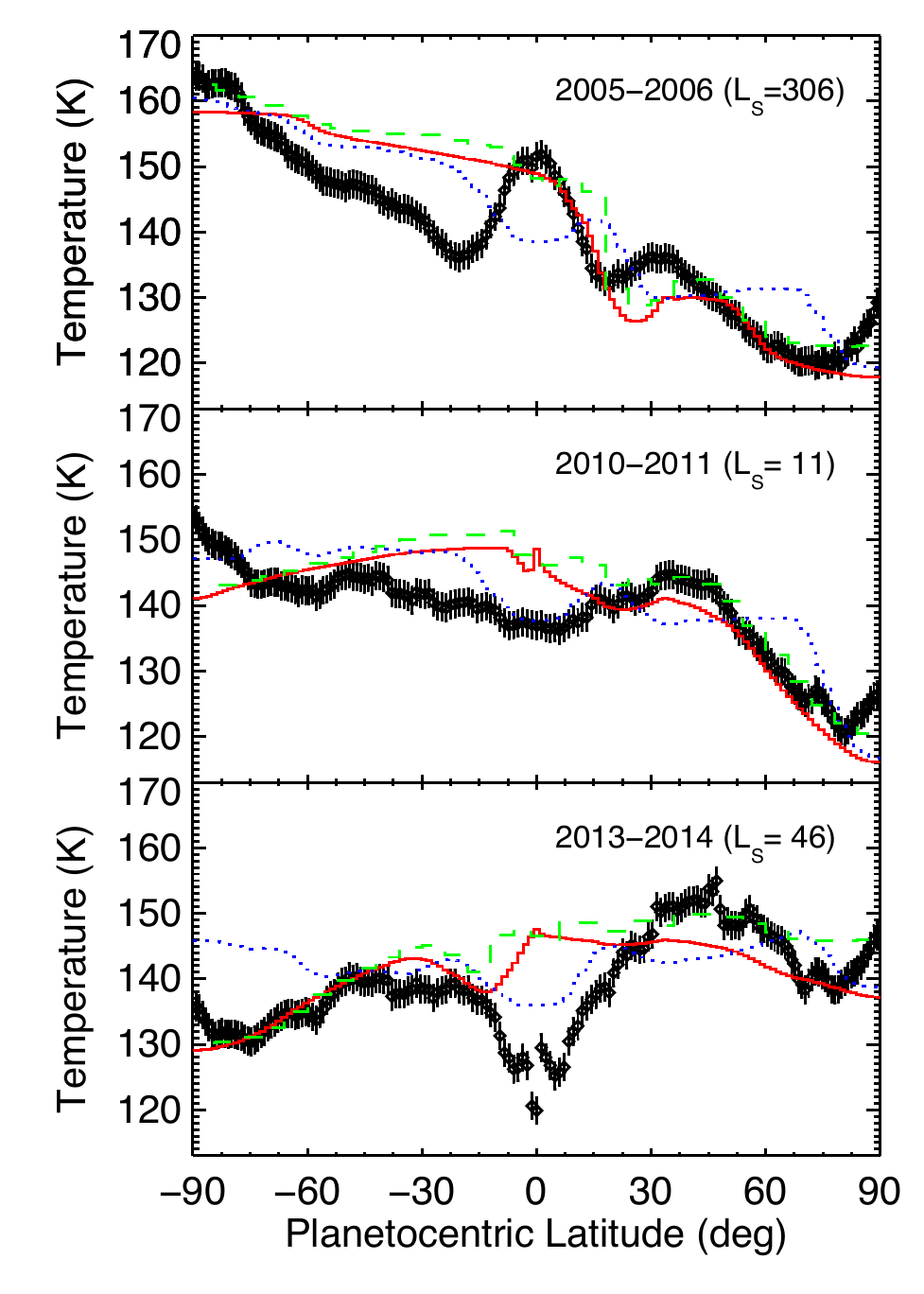}
\caption{Saturn's meridional temperature variations at 1 mbar at 3 different epochs derived from nadir Cassini/CIRS observations, compared with model predictions from \citet{14guerlet} as the solid line, \citet{12friedson} as the dotted line, \citet{10greathouse} as the dashed line. }
\label{comp1.8mbar}
\end{center}
\end{figure}

These three modern models are based largely on the same underlying assumptions, and so we compare each to Cassini measurements of the 1-mbar temperature contrasts in Fig. \ref{comp1.8mbar} to illuminate interesting seasonally forced effects and localized dynamical effects.  The TG model assumes the meridional and vertical mixing ratios for C$_2$H$_2$ and C$_2$H$_6$ as measured by \citet{09guerlet}, and assumes that this distribution is fixed with time.  The TG model is the coarsest (in latitude sampling) of the three models, and does not include the effects of dynamics (dashed line in Fig. \ref{comp1.8mbar}).  The FM model uses the latitudinal average of the vertical profiles of C$_2$H$_2$ and C$_2$H$_6$ as measured by \citet{09guerlet} and holds these vertical profiles constant with time and latitude, even though they are accounting for dynamical redistribution of heat via diffusion and advection (dotted line in Fig. \ref{comp1.8mbar}).  The SG model uses the average of the vertical mixing ratio profiles for C$_2$H$_2$ and C$_2$H$_6$ between $40^\circ$N and $40^\circ$S (planetocentric) as taken from \citet{09guerlet} (solid line in Fig. \ref{comp1.8mbar}).  These mean vertical profiles are assumed constant with latitude and time.  None of the models feature the sharp upturns in C$_2$H$_2$ and C$_2$H$_6$ abundances at high latitudes (nor the added contribution of polar stratospheric aerosols), and so are unlikely to reproduce the radiative energy balance near the poles \citep{15fletcher_poles, 15guerlet}.

A comparison between the three models reveals how the different assumptions manifest themselves in the final predicted temperatures and how those final temperatures compare to the measurements.  Given these assumptions, one would expect that the TG and SG models should return very similar results with the differences between the two being due to the different C$_2$H$_2$ and C$_2$H$_6$ distributions assumed, which is in fact what we see in Fig. \ref{comp1.8mbar}.  Similarly the comparison between the FM and the SG/TG models show the effects of dynamical diffusion and advection of heat on temperatures, as the radiative heating and cooling scheme in all three models are quite similar.  Where the FM model (dotted) is significantly cooler/hotter than the SG model (solid), this could be related to upwelling/downwelling in the FM model causing adiabatic expansion/compression.  However, this assumes identical treatment of radiative balance between the three models, and neglects other sources of heating and cooling (e.g., interactions between waves and the mean flow).

As can be seen in Fig. \ref{comp1.8mbar}, all three models do an impressive job of reproducing the main pole to pole temperature variations as measured by Cassini/CIRS nadir observations \citep{10fletcher_seasons, 15fletcher_poles}.  The strong southern summer temperature gradient seen in the top panel of Fig. \ref{comp1.8mbar} ($L_s=306^\circ$) transitioning to a more equal north/south temperature profile by mid-northern spring ($L_s=46^\circ$) is well tracked by the models, with the exception of the FM model which does not cool as fast at high southern latitudes.  However, several discrepancies stand out.   The temperatures between $\pm20^\circ$ planetocentric latitude are completely different compared to the three model predictions.  Temperatures in this region are significantly altered by the dynamical phenomenon known as Saturn's Semi-Annual Oscillation (SSAO) \citep{08fouchet, 08orton_qxo, 11guerlet}.  The TG and SG models are purely radiative and thus could not possibly match the temperature variation of this dynamical phenomenon.  While the FM model could possibly reproduce such a dynamical event, they suggest that the coarseness of their vertical grid may have made it impossible to resolve the waves needed to force the SSAO.  Another complication is the heating associated with Saturn's stratospheric vortex \citep{12fletcher}.  The nominal seasonal trend of stratospheric temperatures at mid-northern latitudes was disrupted by the production of the stratospheric vortex (known as the `beacon', see Chapter 13) in late 2010, and although this region was avoided to produce the measured temperatures in  Figure \ref{comp1.8mbar}, the storm-induced heating of the northern mid-latitudes is the likely cause for the mismatch with the model predictions in 2013.

One of the starkest discrepancies between the models and the data in Fig. \ref{comp1.8mbar} appears at the polar latitudes.  The 1-mbar temperatures measured by Cassini at the northern and southern poles are compared to the three models in Fig. \ref{polartemp}, to assess their capabilities for reproducing the absolute temperatures and the timing of the polar minima/maxima in temperatures.  Although the timing of the south polar maximum is reproduced in the models (1-2 years after summer solstice), the south polar temperatures are elevated over the predictions of all three models, and the temperature range is larger in the data than in the model predictions.  At the north pole, the models are unable to reproduce the observation that the coldest temperatures are identified between $75-80^\circ$N, rather than at the north pole itself, although the 6-8 year phase lag between the winter solstice and the coldest temperatures is consistent with the models.  It is suspected that these differences are due to circulations associated with polar vortices, or the increasing importance of stratospheric aerosols or enhanced hydrocarbons in the radiative budget at high latitudes \citep{15fletcher_poles}. Interestingly, the FM model stays warmer for longer at the south pole, over-predicting the southern temperatures in 2013-14.  This large temperature increase poleward of $60^\circ$S seen in the FM model is the result of extensive downwelling of gas, whereas observations show that this region has in fact cooled substantially for the duration of Cassini's observations (Fig. \ref{polartemp}).  While the downward advection of material naturally heats the gas by adiabatic compression, one would also expect an increase in cooling due to an increase in C$_2$H$_2$ and C$_2$H$_6$ mixing ratio.  If advection of material to serve as coolants (i.e., C$_2$H$_2$ and C$_2$H$_6$ concentrations, stratospheric aerosols) could be included in the FM model, it might help explain the mismatch between the data and model.  

\begin{figure}%
\begin{center}
\figurebox{3.2in}{}{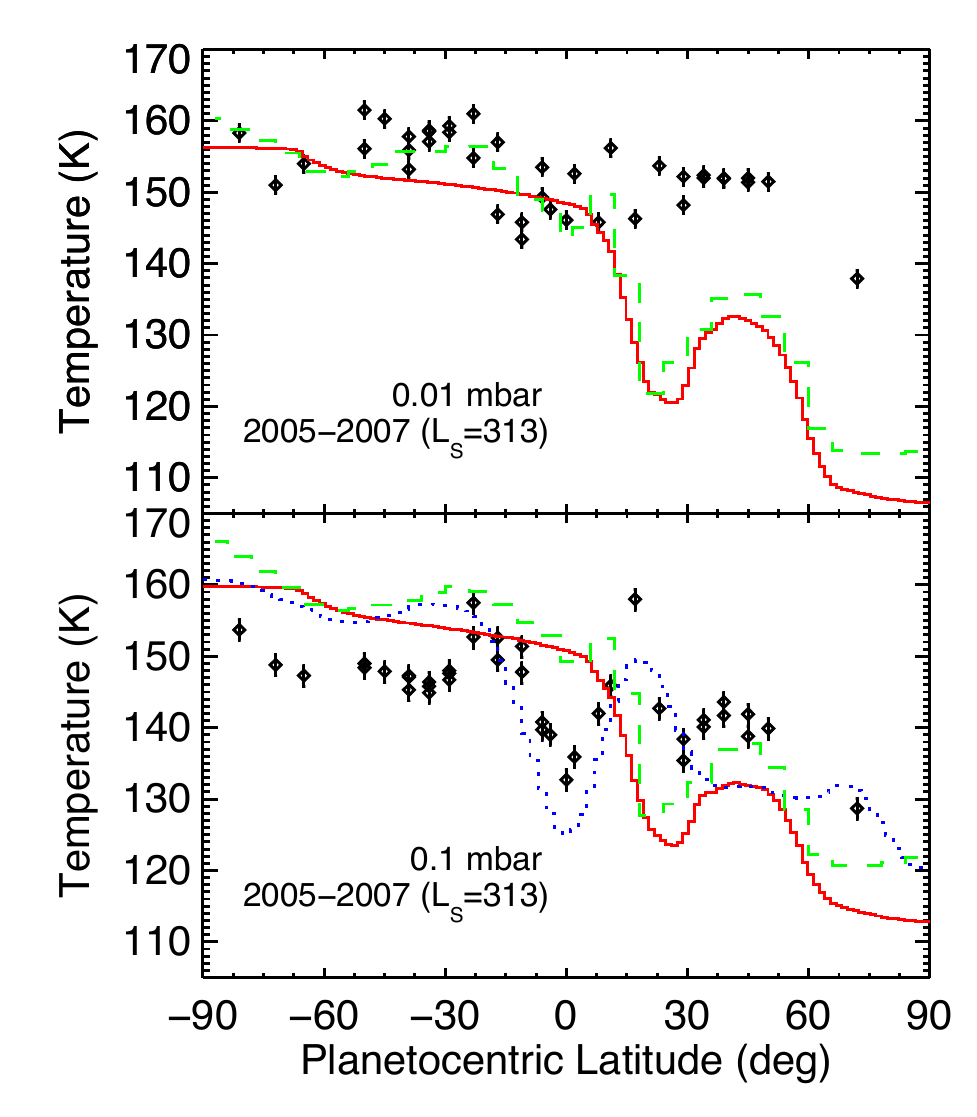}
\caption{Model-data comparisons for temperatures at the 0.01 and 0.1 mbar level.  The observed temperatures were retrieved using Cassini/CIRS limb observations in late southern summer \citet{09guerlet}.  Models are identified as the solid line for \citet{14guerlet}, the dotted line for \citet{12friedson}, and the dashed line for \citet{10greathouse}.  Note the Friedson \& Moses model is not shown at the 0.01 mbar level as it does not accurately model Saturn's atmosphere at that pressure level.}
\label{complowp}
\end{center}
\end{figure}

We now return to the curious fact that the seasonal response measured at higher stratospheric altitudes using CIRS limb observations (0.01-0.1 mbar) by \citet{09guerlet} and \citet{15sylvestre} appears to be smaller than that at 1 mbar.  Fig. \ref{complowp} shows the comparison of the three models to the limb observations in southern summer ($L_s=313^\circ$) at 0.1 and 0.01 mbar.  Although there is remarkable agreement with the temperatures in southern summer (particularly at 0.01 mbar), the measured temperatures are much higher than model predictions in the north.  This could be the result of several processes - near $25^\circ$N the effect of ring shadowing may have been overestimated, or significant mixing/advection might wash out the ring shadow effects seen in the purely radiative models.  Although the FM model does not offer useful results at the 0.01-mbar level, it does at the 0.1-mbar level.  There we can see that the cool region of temperatures due to the ring shadow (15-30$^\circ$N) seen in the TG and SG models is warmed significantly by subsidence in the FM model.  It is likely this effect occurs at 0.01 mbar as well, and may help to explain the model-data mismatch at the higher altitudes.  Away from the ring-shadowed region, additional sources of upper atmospheric heating, such as gravity wave breaking or from some interaction with the thermosphere, might be the culprit for the warmer temperatures over the rest of the northern hemisphere.  

\begin{figure}%
\begin{center}
\figurebox{3.2in}{}{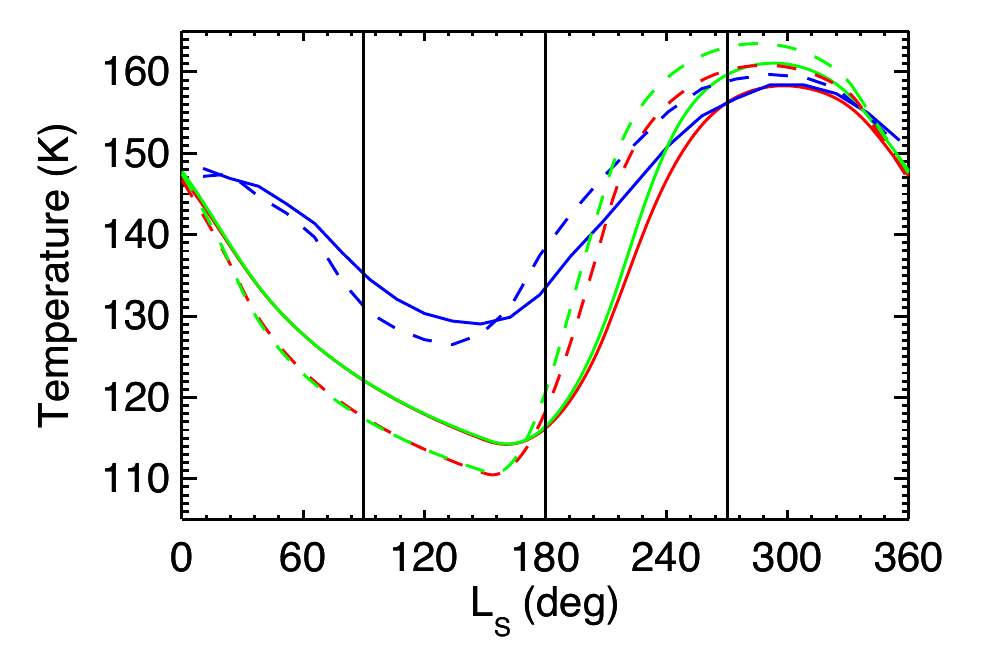}
\caption{Model/model comparison at $75^\circ$S between the \citet{14guerlet} model in red, \citet{12friedson} model in blue and the \citet{10greathouse} model in green.  The solid lines represent the change in temperature over time at the 1.0-mbar level while the dashed lines are for the 0.1-mbar level.  The vertical black lines represent the times of Northern Summer Solstice, Northern Autumnal Equinox, and Northern Winter Solstice, from left to right respectively.}
\label{modelcomp}
\end{center}
\end{figure}

Figs.  \ref{polartemp}, \ref{comp1.8mbar} and \ref{complowp} demonstrate how well the models perform relative to the measured data and at specific instances in time.  It is also useful to compare how the models behave over an entire Saturn year at a single latitude to learn about seasonal phase lags.  In Fig. \ref{modelcomp} we plot the predicted temperatures of Saturn from all three models at $75^\circ$S planetocentric latitude.  The solid lines represent the temperatures at 1 mbar while the dashed are from 0.1 mbar.  All of the models show that the peak temperature in southern summer occurs about $30^\circ$ of $L_s$ after the southern summer solstice ($L_s=270^\circ$) at 1 mbar, but only about $15^\circ$ of $L_s$ at 0.1 mbar.  This is due to the lower-pressure level having less mass, and thus less thermal inertia, in addition to the fact that the atmospheric coolants are more abundant at lower pressures.  The models all suggest that the maximum and minimum in temperatures are more extreme and can change more rapidly at lower pressures due to the lower thermal inertia found there, which appears to be oddly inconsistent with the limb-data analyses from Cassini.  Finally, the model of FM is significantly different from those of TG and SG.  The reduction of the wintertime temperatures during polar night is not surprising as the purely radiative seasonal models continually radiate to space whilst in darkness without accounting for any heat advection, but it is probable that significant diffusion and advection of heat would occur in this region.  However, it is interesting to note that the models that do the best job of reproducing the measured temperatures at $L_s=46^\circ$ (i.e., southern autumn, Fig. \ref{seasonal_temp}) and $75^\circ$S latitude are those without dynamics.  

In summary, it appears that the dominant driver of Saturn's stratospheric temperatures is solar forcing, albeit perturbed by circulation and dynamics.  While purely radiative seasonal models do not reproduce the measured temperatures perfectly, they seem to do as good a job as the more complicated dynamical model of FM.  Advection of the hydrocarbon abundances along with the heat may be significant, in addition to the seasonally shifting contribution of tropospheric and stratospheric aerosols and other heating sources (e.g., wave activity) not currently accounted for in the theoretical models.  Hopefully, with the advent of new 3D GCMs with proper radiative forcing and higher spatial resolution (required for resolving waves and instabilities that influence the circulation), a model will be produced in the near future with a self consistent radiative, chemical and dynamical system.

\subsection{Implications of seasonal temperature change}
\label{dynamics}

The seasonal shifting of hemispheric temperature asymmetries has implications for both Saturn's global energy budget and atmospheric circulation (e.g., polar atmospheric circulations, wave phenomena, tropospheric overturning, storm perturbations, Hadley circulations).  \citet{10li} demonstrated that the shifting thermal asymmetries measured by CIRS had implications for Saturn's emitted power (average of $4.952\pm0.035$ W/m$^2$), which decreased by 2\% from 2005 to 2009 and was found to be larger in the south ($5.331\pm0.058$ W/m$^2$) than in the north ($4.573\pm0.014$ W/m$^2$).  The peak contribution to the outgoing emission (near 320 mbar) was found to be shallower in the northern winter hemisphere than in southern summer, potentially due to seasonal asymmetries in aerosol content in the upper troposphere, or the details of the upper tropospheric temperature structure.  Thus emitted power measurements over a full Saturn year are desirable to properly constrain the energy budget.

Meridional temperature gradients are related to the vertical shear on the zonal winds via the thermal wind equation, as the Coriolis forces on the winds should remain in geostrophic balance with the horizontal pressure gradients.  The stratospheric $dT/dy$ is indicative of positive vertical shear in the winter and negative vertical shear in the summer \citep{12friedson}.  Cassini's seasonal monitoring has shown that $dT/dy$ has become increasingly positive in the mid-stratosphere over the ten years of the mission (see Fig. \ref{temp_profile}), meaning that northern middle-atmospheric prograde jets should become more retrograde (i.e., decreasing eastward velocities) and southern prograde jets should become more prograde (increasing eastward velocities).  These modifications to the stratospheric wind field, although inferred indirectly from temperatures, can have implications for the transmissivity of waves upwards from the convective troposphere \citep[e.g., the wave transport of energy from Saturn's storm regions in 2010-11,][]{12fletcher}.  Despite the large temperature fluctuations, thermal wind variability in the troposphere has been small over the duration of the Cassini mission \citep[variations smaller than 10 m/s per scale height at the tropopause,][]{15fletcher_fp1}, with the largest changes in the northern flank of the equatorial jet.

Finally, although radiative climate models successfully reproduce the magnitude and scale of the observed asymmetries, they lack the dynamic perturbations that govern the temperatures on a smaller scale.  When purely radiative calculations are insufficient to reproduce the observed temperature changes, we can use the thermodynamic energy relationship \citep{03hanel, 04holton} to relate the change in the temperature field to the residual mean circulation causing net heating (subsidence) and cooling (upwelling).  \citet{90conrath} used these techniques to show that a diffuse inter-hemispheric circulation was expected at solstice, with rising motion in the summer hemisphere and downward motion in the winter hemisphere.  However, at equinox the flow consisted of two cells, with rising motion at low latitudes and subsidence poleward of $\pm30^\circ$ latitude, which does not appear to correspond to the equinoctial observations of Cassini.  \citet{12friedson} produced a more complex model to predict these inter hemispheric transports, finding a seasonally-reversing Hadley-like circulation at tropical latitudes and cross-equatorial flow from the summer into the winter hemisphere twice per year.  \citet{15fletcher_poles} used the deviation of Saturn's measured temperatures from radiative equilibrium to show that upwelling/downwelling winds with zonally-averaged vertical velocities of $|w|\approx0.1$ mm/s could account for the cooling/warming of the southern/northern polar stratospheres at 1 mbar, respectively.  In summary, the general warming of the northern hemisphere and cooling of the southern hemisphere over the duration of the mission could be reproduced by an inter-hemispheric transport into the spring hemisphere in the stratosphere and upper troposphere, perturbed at low latitudes by a Hadley-type circulation, and at high latitudes by the formation and dissipation of polar vortices.

\subsection{Non-seasonal phenomena}

\begin{figure}%
\begin{center}
\figurebox{3.2in}{}{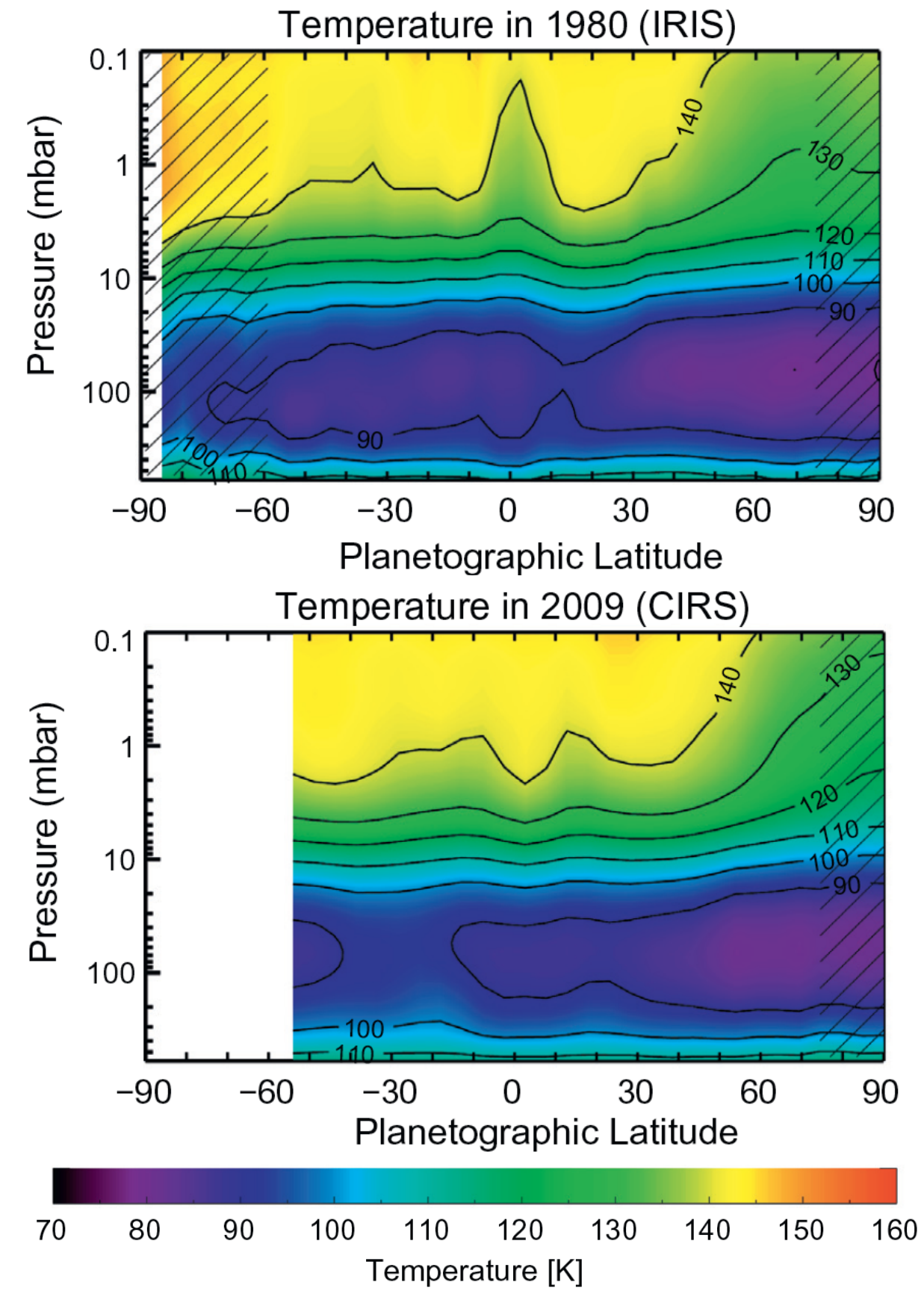}
\caption{Zonal mean temperature retrievals from Voyager/IRIS ($L_s=8^\circ$) and Cassini/CIRS ($L_s=3^\circ$) as presented by \citet{14sinclair}.  Although these differ by $\Delta L_s=5^\circ$, the comparison shows that the zonal mean stratospheric temperatures differ considerably at the equator between the two epochs.  The upper troposphere also appears to be warmer in the southern autumn hemisphere (10-45$^\circ$S) in the Cassini measurements compared to the Voyager measurements, as discussed by \citet{13li}, suggesting non-seasonal variability.   The hashed areas are regions of low retrieval confidence.}
\label{voyagertemp}
\end{center}
\end{figure}

At low latitudes, the signature of Saturn's evolving seasons appear to be overwhelmed by the phenomenon known (perhaps incorrectly) as Saturn's Semi-Annual Oscillation (SSAO).  Originally identified by \citet{08fouchet} and \citet{08orton_qxo} as temporally-evolving oscillations in the vertical and latitudinal temperature structure, the downward propagation of the wave structure has been studied by Cassini infrared spectroscopy and radio occultation data \citep{10fletcher_seasons, 11guerlet, 11li, 11schinder}.  These studies revealed that the local temperature extrema observed in 2005 (an equatorial local maximum located at the 1-mbar pressure level and a local minimum located at 0.1 mbar) descended by approximately 1.3 scale heights in 4.2 years (see Fig. \ref{seasonal_temp}). The downward propagation of the oscillation was consistent with it being driven by absorption of upwardly propagating waves and suggested a $\sim$15-year period for Saturn's equatorial oscillation, as already derived from long-term ground based observations \citet{08orton_qxo}.  

Cassini observations in early northern spring might be expected to replicate Voyager 1 and 2 observations taken one Saturnian year earlier ($L_s=8-18^\circ$).  These datasets have been compared by \citet{13li}, \citet{14sinclair} and \citet{15fletcher_fp1}, and mid-latitude temperatures were largely the same between the two epochs, confirming a repeatability of the broad seasonal cycle. In the stratosphere, \citet{14sinclair} showed that CIRS 2009-10 temperatures were $7.1\pm1.2$ K warmer than IRIS 1980 temperatures near 2 mbar in the equatorial region (Fig. \ref{voyagertemp}), implying that Voyager and Cassini are capturing the equatorial oscillation in slightly different phases, which is inconsistent with the previously-identified period of the SSAO.  Indeed, this suggests that the oscillation may be quasi-periodic, as on Earth.  It is possible that seasonal storm activity, such as the equatorial eruption in 1990 and the northern mid-latitude eruption of 2010 (see Chapter 13), may have disrupted this equatorial oscillation.    

Evidence for non-seasonal variability was also observed in the troposphere \citep{13li, 15fletcher_fp1}, potentially as a result of the wave structure impinging on the stably-stratified upper troposphere.  \citet{13li} suggested that CIRS results were 2-4 K warmer than IRIS results at $\pm15^\circ$ latitude near 300 mbar, and that the tropopause (50-100 mbar) is 8-10 K warmer in 2009/10 compared to 1980.  Although low-pressure ($p<100$ mbar) changes were not confirmed by \citet{14sinclair} and \citet{15fletcher_fp1}, they do confirm the changing latitudinal thermal gradients near the equator (and hence the vertical shear on zonal winds), which were rather different between the Voyager and Cassini epochs.  The distribution of para-H$_2$ also suggested substantial differences in the tropical region between Voyager and Cassini \citep[see Section \ref{trop_chem} and][]{15fletcher_fp1}.  Taking the tropospheric and stratospheric results together, it appears that Saturn's equatorial oscillation may not be strictly `semi-annual' at all, and possibly changes from year to year in response to dynamic phenomena such as storm eruptions. 

\section{Distribution of chemical species}
\label{chem}

\subsection{Overview of Saturn's atmospheric composition}
\begin{figure}%
\begin{center}
\figurebox{3.4in}{}{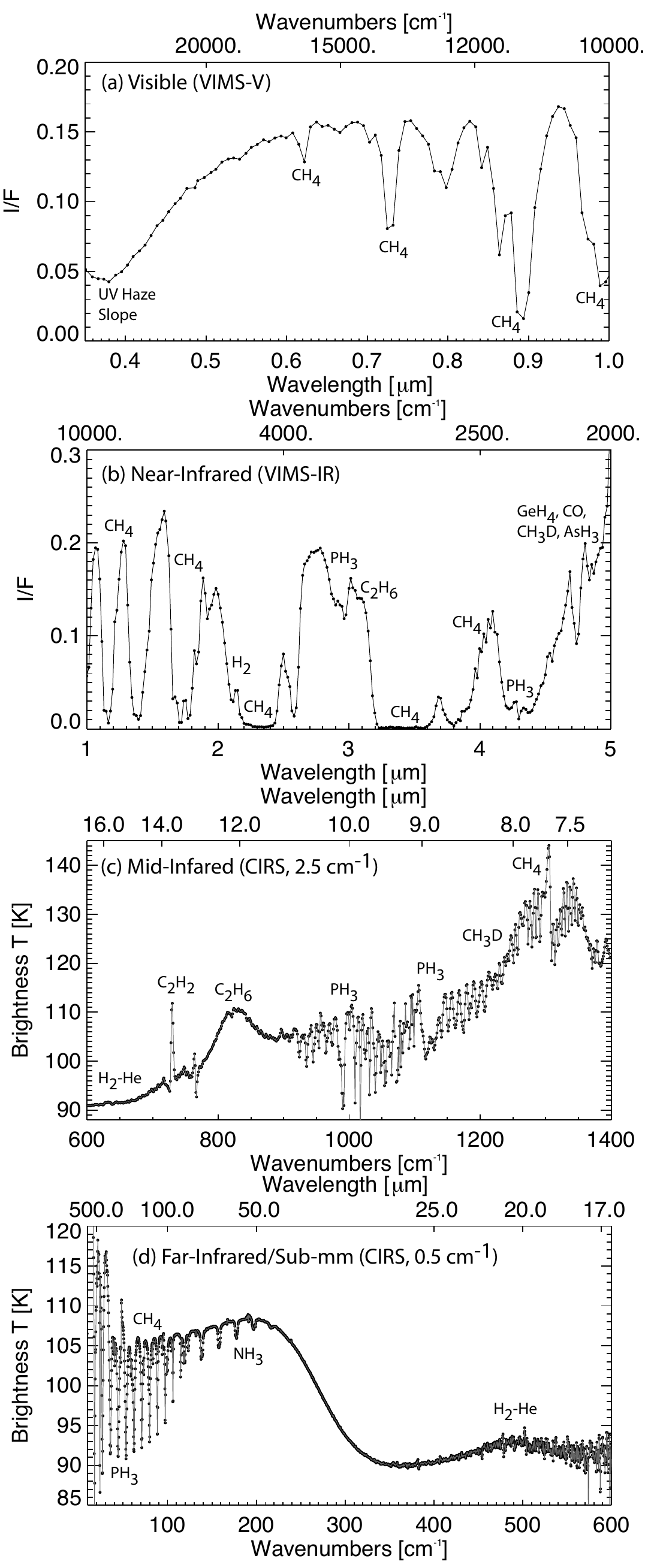}
\caption{Examples of remote sensing data from the UV to the sub-millimeter as measured by Cassini/VIMS (top two panels) and Cassini/CIRS (bottom two panels), demonstrating the rich variety of absorption and emission features.  Note that the VIMS panels are provided in wavelength and reflectivity units, whereas the CIRS panels are provided in wavenumber and brightness temperature units, to maximize the visibility of features.  All spectra were averaged over Saturn's equatorial region, and can be expected to vary from location to location on the planet.  Not shown are Cassini/UVIS observations at the shortest wavelengths (55-190 nm) and Cassini/RADAR observations at the longest wavelengths (2.2 cm).}
\label{sat_spectra}
\end{center}
\end{figure}

The composition and 3D distribution of gas-phase constituents in Saturn's atmosphere is controlled by the coupled influence of thermochemical equilibrium, photochemistry and other disequilibrium chemical mechanisms, global circulation and regional atmospheric dynamics, and aerosol microphysical processes.   In the absence of \textit{in situ} sounding of Saturn's atmosphere (see Chapter 14), we rely on remote sensing measurements (Fig. \ref{sat_spectra}) to determine the spatial distribution of these species, and theoretical chemical models to explain {\em why\/} particular species are observed or not observed.  Saturn's hydrogen-helium atmosphere contains a wealth of reduced trace species deriving from the dominant chemical elements (carbon, nitrogen, sulfur, phosphorus and oxygen), and thermochemical equilibrium in the deep hot atmosphere steers the bulk atmospheric abundances toward the dominant H$_2$, He, H$_2$O, CH$_4$, NH$_3$, and H$_2$S atmospheric composition at the low temperatures and pressures found in the observable upper atmosphere \citep[e.g.,][]{85fegley}.  Metals, silicates, and other refractory species condense out deep in the atmosphere and remain unobservable.  Nitrogen, sulfur and oxygen compounds are largely hidden from the reach of remote sensing due to condensation into cloud decks (Section \ref{clouds}), with the exception of NH$_3$, which is volatile enough that its presence can be detected in the upper troposphere within its condensation region. Oxygen compounds can be detected in the upper atmosphere when delivered from sources external to Saturn.  Temperatures at Saturn's tropopause remain too warm for methane condensation, so that this principal carbon-bearing molecule remains abundant throughout the observable upper troposphere and stratosphere.

Chemical equilibrium is not always maintained, however, as the gases at depth are convected upward into cooler regions where chemical-kinetic reactions can become more sluggish.  Species mixing ratios ``quench'' when vertical transport time scales fall below the chemical kinetic time scales for conversion between different major forms of an element \citep[e.g.,][]{77prinn,84lewis}.  Such quenching is responsible for the presence of CO, PH$_3$, GeH$_4$, and AsH$_3$ in Saturn's upper troposphere, whereas equilibrium arguments suggest that phosphorus, germanium and arsenic should be sequestered in condensates at depth \citep[e.g.,][]{94fegley}.

\begin{figure}%
\figurebox{3.2in}{}{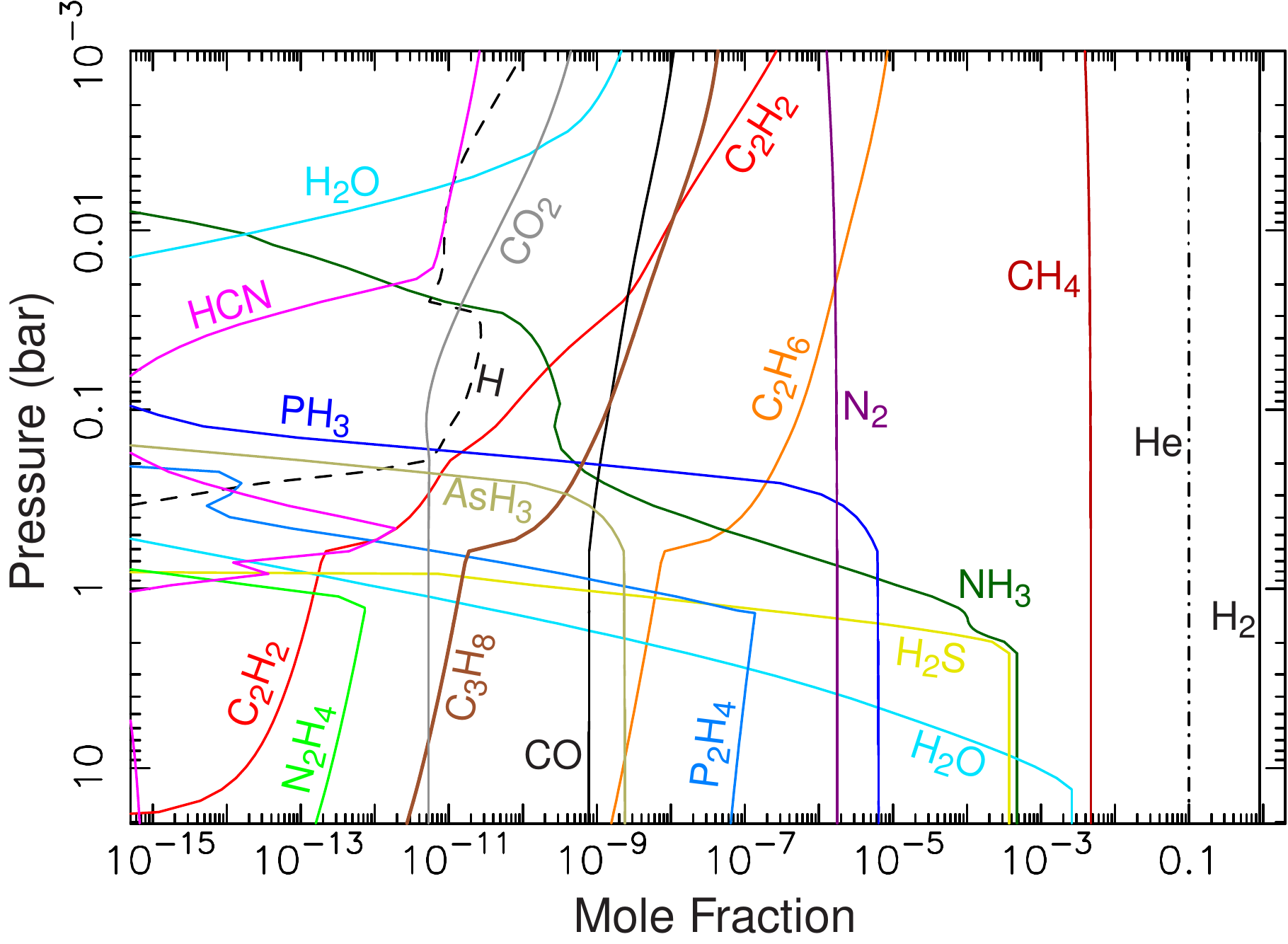}
\caption{Predicted vertical distributions of several tropospheric species of interest on Saturn for $27^\circ$S planetographic latitude under summer conditions, based on the photochemical model described in \citet{10moses}.  The sharply decreasing abundance of H$_2$O, H$_2$S, NH$_3$, P$_2$H$_4$, and N$_2$H$_4$ with increasing altitude in the 0.1-10 bar region is due to condensation, whereas the sharp decrease in the PH$_3$ abundance near 0.2 bar is due to photochemistry.  Molecules whose mixing ratios increase with increasing altitude have a source at higher altitudes, either from stratospheric hydrocarbon photochemistry (e.g., C$_2$H$_6$, C$_2$H$_2$, C$_3$H$_8$, HCN) or from influx of external material combined with photochemistry (e.g., H$_2$O, CO, CO$_2$).}
\label{figtropchem}
\end{figure}

When a species is able to avoid cold-trapping by condensation, it can be transported to sufficiently high altitudes to interact with solar UV radiation, and can serve as a parent molecule for chains of chemical reactions that produce additional disequilibrium species in the upper troposphere and stratosphere (e.g., N$_2$H$_4$, P$_2$H$_4$ and complex hydrocarbons).  Saturn's axial tilt and seasonally variable solar insolation as a function of latitude result in temporal and meridional variations in the photochemical production and loss rate of atmospheric species.  This variable photochemistry, combined with atmospheric dynamics, controls the distribution of the atmospheric constituents in the stratosphere and upper troposphere, as shown in Fig. \ref{figtropchem}.  In this section, we review our current knowledge of Saturn's seasonally-variable composition, building upon the extensive reviews by \citet{84prinn} and \citet{09fouchet}, and showing how the species distributions are intricately connected with Saturn's radiative budget (Section \ref{temp}) and aerosols (Section \ref{clouds}).  We briefly describe the dominant processes and discuss the advances in our understanding of atmospheric chemistry that have occurred since the review by \citet{09fouchet}.  We confine our discussion to Saturn's troposphere and stratosphere; the thermosphere and ionosphere are covered in Chapter 9.   

\subsection{Tropospheric composition}
\label{trop_chem}

\subsubsection{Condensible volatiles}
\label{nh3}

The vertical structure of Saturn's cloud-forming region is determined by thermochemical equilibrium, condensation chemistry and vertical mixing of quenched species.  Thermochemical equilibrium models for Saturn have been developed by \citet{85fegley}, \citet{94fegley}, \citet{02lodders}, \citet{05visscher} and \citet{06visscher}. Water, the dominant equilibrium form of oxygen on Saturn, will condense in the upper troposphere to form a liquid aqueous solution cloud at depth (around 20 bar, although local temperatures, meteorology and the bulk abundances will affect these cloud condensation levels), trending to water ice at higher altitudes \citep[e.g.,][]{73weidenschilling}.  The pressure at the cloud base will depend on the unknown bulk oxygen abundance on Saturn, as well as on the details of regional atmospheric dynamics \citep[e.g.,][]{11sugiyama, 14sugiyama, 14palotai}.  Because some of the planet's oxygen is tied up in condensed silicates at even deeper levels, the water abundance at the base of the aqueous solution cloud represents an already depleted fraction of the bulk oxygen abundance.  \citet{05visscher} determine that roughly 20\% of Saturn's bulk oxygen will be tied up in these condensates if the oxygen-to-silicate-and-metal fraction in the atmosphere is in solar proportions.  Indeed, signatures of Saturn's tropospheric water have only been detected near 5 $\mu$m by ISO \citep[probing the 2-4 bar level above the cloud,][]{97degraauw}, and the mixing ratios were highly sub-solar.  Cassini/VIMS spectra of the same region were unable to provide sufficient sensitivity to measure the water abundance \citep{11fletcher_vims}. 

Ammonia and hydrogen sulfide are the dominant equilibrium forms of nitrogen and sulfur, respectively, in Saturn's atmosphere.  Some NH$_3$ will be dissolved in the upper-tropospheric water solution cloud.  Above that level, gas-phase NH$_3$ and H$_2$S are expected to react to form a crystalline NH$_4$SH cloud, followed at even higher altitudes by a cloud of NH$_3$ ice.  \citet{89briggs} used 1.3-70 cm radio observations from the VLA and Arecibo in 1980 to measure a depletion in NH$_3$ from 25 bar to 2 bar, suggesting that the formation of the NH$_4$SH cloud was the main NH$_3$ sink, which indirectly required H$_2$S to be ten times solar composition \citep[approximately 400 ppm, following][but depending on the solar sulfur abundances chosen]{99vandertak}.  However, uncertainties in the spectral line shapes and data calibration in the radio region, combined with the relatively featureless spectra of the various constituents, make this result highly uncertain, and H$_2$S has never been directly detected.  The deep formation of the H$_2$O and NH$_4$SH clouds means that Saturn's bulk oxygen and sulfur abundances therefore remain unknown, and that H$_2$O at least is unlikely to contribute significantly to seasonally-variable photochemistry in the upper troposphere.

Ammonia, on the other hand, is volatile enough to maintain a presence at the uppermost tropospheric altitudes (see Fig.~\ref{figtropchem}).  Signatures of NH$_3$ were first tentatively identified at visible wavelengths by \citet{66giver} and positively identified by \citet{74encrenaz}, and has since been identified in the near-infrared \citep{80larson, 83fink}, far-infrared \citep{84courtin} and microwave wavelengths \citep[the 1.2-cm inversion band,][]{85depater}.  Measurements at infrared wavelengths have provided globally-averaged abundances at a range of altitudes:  microwave and radio observations indicate mole fractions exceeding 500 ppm at $p>3$ bar \citep{85depater, 89briggs, 99vandertak}, declining to abundances of around 100 ppm near the condensation altitude \citep{84courtin, 89grossman, 89briggs, 90noll, 97degraauw, 00orton, 04burgdorf}, and becoming strongly sub-saturated above the clouds \citep[approximately 0.1 ppm at 500 mbar,][]{97kerola, 06kim, 09fletcher_ph3}.  See \citet{09orton} for a historical overview.  These globally-averaged values suggest that ammonia decreases with altitude above the clouds due to condensation \citep[relative humidities of approximately 50\%][]{97degraauw} and photochemical processes.  Both of these processes depend on temperature and seasonal insolation cycles, so we might expect ammonia to be spatially and temporally variable. 

\begin{figure}%
\begin{center}
\figurebox{3.2in}{}{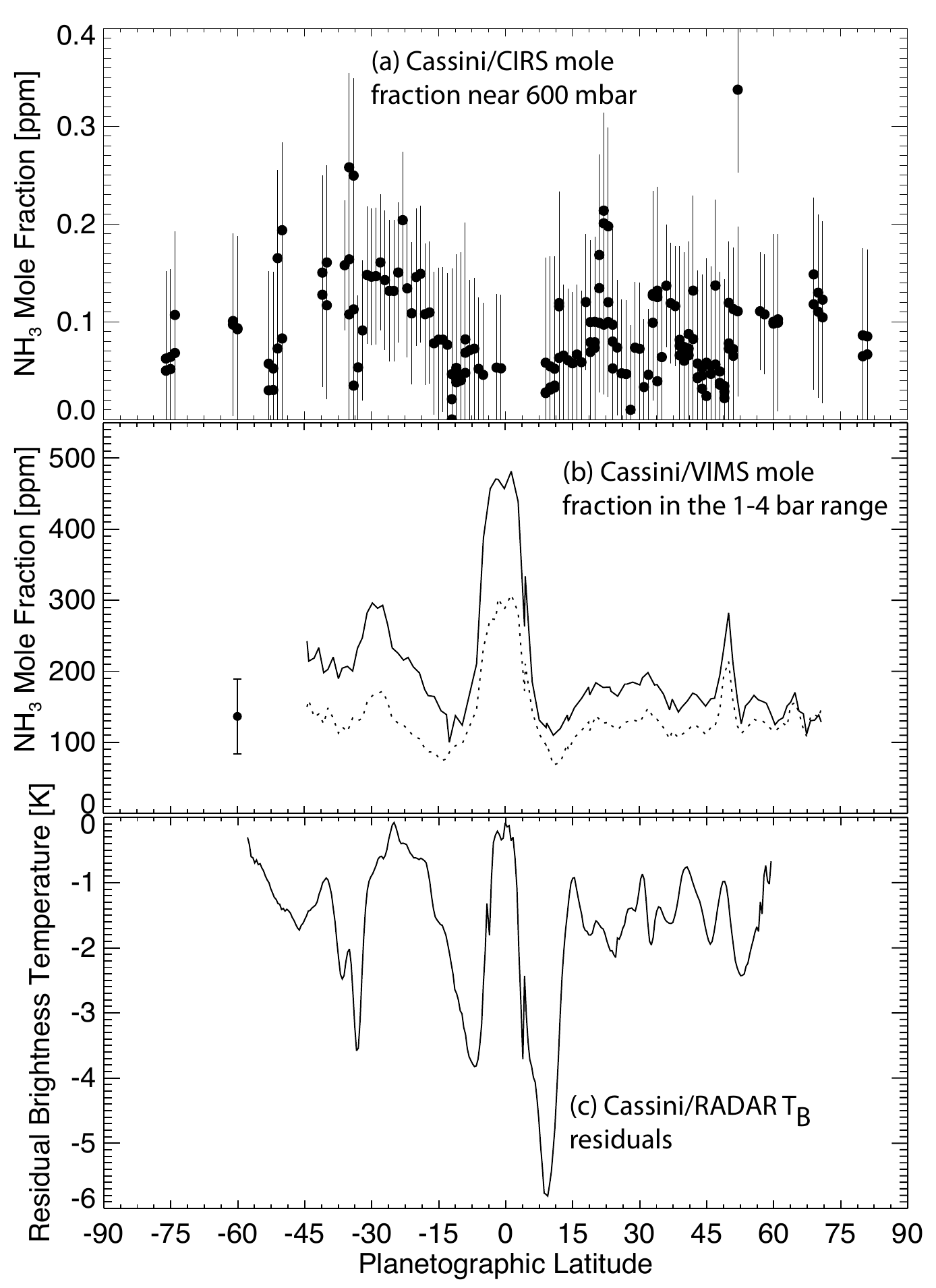}
\caption{Zonal mean ammonia distributions measured (a) in the upper troposphere by Cassini/CIRS \citep{12hurley} and in the cloud-forming region by (b) Cassini/VIMS \citep{11fletcher_vims} and (c) Cassini/RADAR \citep{13laraia}.  The solid line in panel (b) represents a VIMS retrieval in the absence of aerosol scattering, the dotted line shows the effects of scattering.  The isolated point with error bars at $60^\circ$S shows the typical uncertainty on the VIMS inversion.  In panel (c), large negative values of the 2.2-cm brightness temperature residuals imply a low NH$_3$ humidity, and these regions are aligned with regions of depleted ammonia vapor in the VIMS results.  The region between $\pm4^\circ$ is from a single RADAR scan, whereas the rest is an average over five RADAR maps.}
\label{ammonia}
\end{center}
\end{figure}

Spatially-resolved distributions of NH$_3$ have been provided by the CIRS \citep{12hurley}, VIMS \citep{11fletcher_vims} and RADAR instruments \citep{13janssen, 13laraia} on the Cassini spacecraft (Fig. \ref{ammonia}).  \citet{12hurley} modelled CIRS far-IR rotational NH$_3$ lines to determine the 600-mbar NH$_3$ distribution (i.e., above the clouds), finding that NH$_3$ was largely uncorrelated with the belt/zone structure of the thermal field, and instead showed a weak hemispheric asymmetry (with higher abundances in the southern summer hemisphere) and no evidence for peak equatorial abundances (Fig. \ref{ammonia}a).  They concluded that condensation and photolytic processes were shaping the distribution, and that NH$_3$ does not trace the local circulation in this region.  \citet{11fletcher_vims} used VIMS observations of the 5-$\mu$m region to identify a narrow equatorial peak (within $5^\circ$ of the equator) in NH$_3$ at $p>1$ bar in the cloud-forming region, with peak values of 500 ppm similar to the deep abundances observed in the microwave \citep{85depater}.  Elsewhere the NH$_3$ abundances fell to 100-200 ppm in the 1-4 bar region, consistent with previous studies of the condensation region, and there was no indication of a strong hemispheric asymmetry (Fig. \ref{ammonia}b).  2.2-cm brightness temperature maps measured by RADAR \citep{13janssen, 13laraia} were used to study NH$_3$ humidity near 1.5 bar, finding low brightnesses near the equator (consistent with the narrow NH$_3$ peak observed by VIMS) that were flanked by regions of high brightness near $\pm9^\circ$ latitude (Fig. \ref{ammonia}c), and a subtle asymmetry between the northern and southern hemisphere \citep{13janssen}.  Furthermore, \citet{13janssen} demonstrated a remarkable stability in the latitudinal distribution of NH$_3$ measured by RADAR over the 2005-2011 time span of the Cassini data, in contrast to the time-variability suggested by VLA data in the 1980s and 1990s by \citet{99vandertak}.  This raises the possibility of long-term variability in Saturn's banded structure similar to Jupiter's `global upheavals'.   However, during the Cassini epoch, the equatorial maximum in NH$_3$ observed by both VIMS and RADAR, flanked by local minima near $\pm9^\circ$, may be consistent with strong equatorial upwelling flanked by regions of subsidence.  \citet{13laraia} deduced an average humidity of $70\pm15$\% in the cloud region, and showed that their data were consistent with a $3-4\times$ solar enrichment (360-480 ppm) at depth, but depleted for $p<2$ bar.  Fig. \ref{ammonia}c may show a slight hemispheric asymmetry, with higher brightnesses (lower NH$_3$ abundances) in the northern, winter hemisphere, qualitatively consistent with the asymmetry in the upper troposphere identified by \citet{12hurley}.  The possible mechanisms creating this asymmetry in Saturn's southern summer are discussed later in this section.

\subsubsection{Quenching and disequilibrium}
\label{ph3}

\textbf{Phosphine:}
Thermochemical equilibrium arguments predict the absence of several species in Saturn's upper troposphere (phosphine, arsine, germane and carbon monoxide) due to sequestration in the hotter deep interior.  For example, equilibrium models suggest that phosphorus would be tied up in PH$_3$ at great depths on Saturn, but reactions with water should convert the phosphine to P$_4$O$_6$ at temperatures below $\sim$900 K, with condensed NH$_4$H$_2$PO$_4$ becoming the dominant phase at temperatures below $\sim$400 K \citep{85fegley}.  Nevertheless, PH$_3$ has been shown to have a significant effect on Saturn's spectrum, providing a clear indication that phosphorus chemical equilibrium is not achieved on Saturn.  The presence of PH$_3$ is most likely due to rapid vertical mixing, transporting PH$_3$ upwards with a dynamical timescale shorter than the chemical depletion timescale, so that the observed abundances are representative of the ``quenched'' equilibrium conditions at deeper atmospheric levels \citep{84prinn, 94fegley}.  Furthermore, the PH$_3$ profile will be very sensitive to vertical winds, with downwelling winds suppressing the PH$_3$ abundance at the $\sim$100-400 mbar level and upwelling winds enhancing it, making this species a useful tracer of tropospheric dynamics.

Before the arrival of Cassini, PH$_3$ studies focused on disk-integrated abundances and the vertical distribution (evidence for PH$_3$ photolysis).  Following its first detection near 10 $\mu$m \citep{74gillett, 75bregman}, PH$_3$ has been studied in the mid-IR from ground- and space-based observatories \citep{80tokunaga, 84courtin, 97degraauw, 01lellouch}; in the 5-$\mu$m opacity window \citep{80larson, 91noll, 97degraauw}; the 3-$\mu$m reflected component \citep{80larson, 97kerola, 05kim, 06kim}; the sub-millimeter  \citep{94weisstein, 96davis,  00orton, 01orton, 04burgdorf, 12fletcher_spire} and the ultraviolet \citep{83winkelstein, 97edgington}.  Vertical distributions were presented by \citet{91noll, 97degraauw, 01orton, 01lellouch, 06kim, 12fletcher_spire} to show that PH$_3$ declined with altitude above the 500-700-mbar level and was not present in the stratosphere, but none of these studies provided assessments of the spatial distribution, which would be required to understand seasonal contrasts and regions of powerful convective uplift.

\begin{figure}%
\begin{center}
\figurebox{3.2in}{}{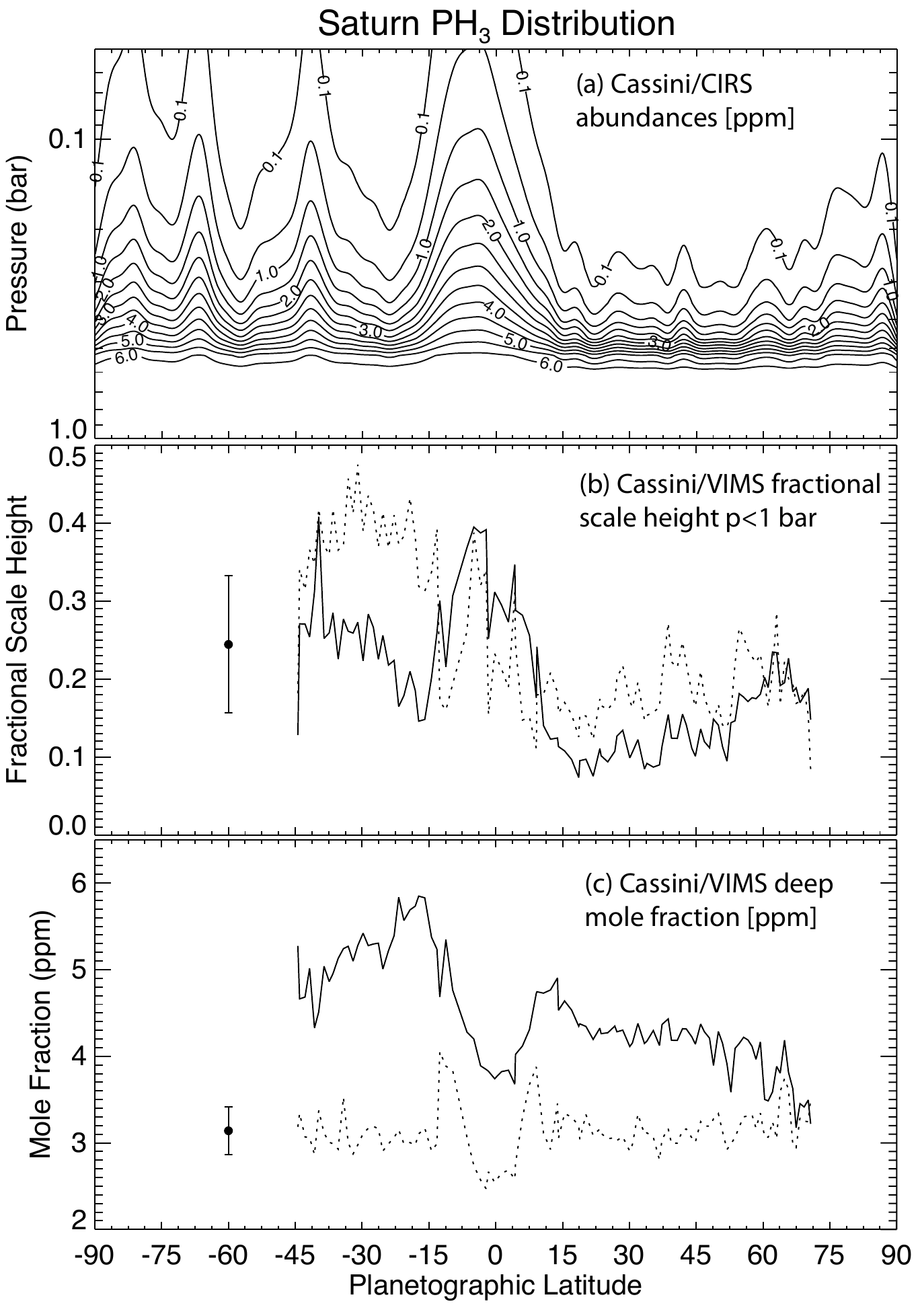}
\caption{Distribution of PH$_3$ determined by Cassini/CIRS \citep{09fletcher_ph3} and Cassini/VIMS \citep{11fletcher_vims} during southern summer, sensing different regions of the troposphere. The CIRS distribution in the upper troposphere (shown as a mole fraction contour plot for $p<1$ bar) and VIMS-derived fractional scale height ($p<1$ bar) are consistent with one another and show an equatorial enhancement and a north-south asymmetry in southern summertime (the fractional scale height is the scale height of PH$_3$ gas divided by the bulk gas scale height).  The deep mole fraction derived from VIMS observations shows opposite trends, suggestive of a different circulation regime beneath the topmost clouds.  Solid lines show VIMS night-side calculations without aerosol scattering, dotted lines show VIMS calculations with scattering of thermal photons.  The two points with error bars at $60^\circ$S show the typical uncertainties on the VIMS inversions.}
\label{phosphine}
\end{center}
\end{figure}

To date, Cassini is the only facility to have provided the spatial distribution of PH$_3$ from Cassini/VIMS (5-$\mu$m) and Cassini/CIRS (10 $\mu$m and sub-millimeter), as shown in Fig. \ref{phosphine}.  Although there are quantitative differences between the mole fractions derived from the two instruments, the results agree qualitatively in the upper troposphere.  A key finding was the enhancement in upper tropospheric PH$_3$ in the equatorial region \citep{07fletcher_ph3, 09fletcher_ph3, 11fletcher_vims}, suggestive of rapid vertical mixing consistent with the cold equatorial temperatures (Section \ref{temp}) and elevated hazes (Section \ref{clouds}) observed there.  This PH$_3$-enriched zone is flanked at tropical latitudes ($\pm23^\circ$ planetographic) by regions of depletion, consistent with a Hadley-style circulation.  However, the equatorial enhancement is only identifiable in the upper troposphere, and not in the deeper adiabatic region (i.e., below the radiative-convective boundary), where the PH$_3$ distribution suggests depleted abundances at the equator and enhanced abundances in the neighboring belts \citep[Fig. \ref{phosphine}(c) and][]{11fletcher_vims}.  CIRS does not have the altitudinal sensitivity to probe pressures greater than 1 bar, so this `deep' distribution of PH$_3$ comes solely from inversions of VIMS spectra in the 5-$\mu$m window, where measurements of composition are rather degenerate with assumptions about tropospheric aerosol properties. As a result, this difference in the zonal mean PH$_3$ distribution between the upper and lower troposphere has yet to be explained, but could be related to a transition in the tropospheric circulation above and below the main cloud decks.    In addition, PH$_3$ was found to be depleted within the cyclonic polar vortices identified at both the summer and winter poles \citep{08fletcher_poles}.   Finally, both CIRS and VIMS observations suggest a hemispheric asymmetry in the PH$_3$ abundance, being more abundant in the southern summer hemisphere.  Whether this asymmetry is permanent or will reverse with Saturn's seasons remains to be seen - \citep{10fletcher_seasons} suggested that the zonal mean PH$_3$ distribution measured by CIRS had not changed during the 2004-2009 period.    The possible seasonal nature of this contrast is discussed below.

\textbf{Other disequilibrium species:}
Ground-based observations of Saturn's 5-$\mu$m opacity window in 1986-1989 provided the first detections of germane \citep[GeH$_4$,][]{88noll}, arsine \citep[AsH$_3$,][]{89noll, 90noll, 89bezard} and carbon monoxide \citep[CO,][]{86noll}.  As with PH$_3$, these species were not expected in the upper troposphere on the grounds of thermochemical equilibrium.  Gaseous GeH$_4$ and AsH$_3$ are expected to be the dominant phases of germanium and arsenic at depth ($T>400$ K), to be replaced by condensed Ge and As phases in the upper troposphere \citep[e.g.,][]{94fegley}, but both species are detected in the upper troposphere, and apparently decrease in abundance with altitude \citep{97degraauw, 89bezard}.  Cassini/VIMS data are theoretically sensitive to both species, but only AsH$_3$ could be mapped spatially given the low spectral resolution \citep{11fletcher_vims}.   As with the deep distribution of PH$_3$, the zonal mean AsH$_3$ distribution showed an equatorial minimum flanked by local maxima at $\pm7^\circ$ and a possible north-south asymmetry (the size of which depended on the assumed scattering properties of Saturn's aerosols in the 5-$\mu$m window).  The global mean abundances of AsH$_3$ \citep[$2.2\pm0.3$ ppb with fully-scattering aerosols,][]{11fletcher_vims} were consistent with ground-based determinations of $3.0\pm1.0$ ppb \citep{90noll} and $2.4^{+1.4}_{-1.2}$ ppb \citep{89bezard}.  

Saturn's CO can originate from two sources:  an external source of oxygen-bearing molecules in the upper stratosphere \citep[detected in the sub-millimeter by ][]{09cavalie, 10cavalie}, and an internal source as a quenched disequilibrium species \citep[measured at ppb-levels in the upper troposphere by][]{86noll, 91noll, 09cavalie}.  These measurements are complex, as the internal and external CO distributions are difficult to separate spectrally, meaning that measurements of the intrinsic CO abundance remain highly uncertain.  For the intrinsic, deep-tropospheric source of CO, recent {\em ab\ initio\/} transition-state theory rate-coefficient calculations and numerical modeling by \citet{11moses} and \citet{11visscher} have advanced our understanding of the CH$_4$-CO quench process on hydrogen-rich planets.  These authors find that thermal decomposition of methanol to form CH$_3$ + OH is the rate-limiting step in the CO $\rightarrow$ CH$_4$ conversion process under solar-system giant-planet conditions.  The resulting thermochemical kinetics and transport models can reliably predict the resulting quenched CO abundance in the upper tropospheres of the giant planets, and possibly provide indirect constraints on the deep water abundance and transport rates if the kinetic conversion between CH$_4$ and CO at high temperatures and pressures were well characterized  \citep[e.g.,][]{77prinn,84lewis,88yung,94fegley,02lodders,02bezard,05visscher,10visscher_co}.  The external source of CO (cometary impacts, continuous interaction with the rings and satellites) are discussed in Section \ref{oxygen}.

Theoretical models suggest that N$_2$ is the most abundant quenched thermochemical species on Saturn and the other giant planets, with a mole fraction of $\sim$1 ppm (\citealt{10moses}; see also \citealt{81prinn}, \citealt{84lewis}, \citealt{85fegley}, \citealt{02lodders}), with the rate-limiting step in the N$_2$ $\rightarrow$ NH$_3$ conversion being the reaction of H with N$_2$H$_2$ \citep{10moses,11moses}.  Because it is a homonuclear molecule, N$_2$ is difficult to observe, but there are potential photochemical consequences of a large N$_2$ abundance in the high-altitude stratosphere and ionosphere that may be observable, particularly at UV wavelengths.  Other quenched molecules such as CO$_2$, HCN, CH$_3$NH$_2$, H$_2$CO, CH$_3$OH, CH$_2$NH, and HNCO are predicted to have very small mixing ratios \citep{10visscher_co,10moses} and so far only upper limits on HCN, H$_2$CO and CH$_3$OH have been reported from sub-millimeter spectroscopy \citep{12fletcher_spire}.

\subsubsection{Tropospheric photochemistry}
\label{trop_photochem}

Asymmetries in the zonal mean distributions of upper tropospheric PH$_3$ and NH$_3$ have been identified in Cassini/CIRS, VIMS and RADAR observations, likely related to seasonal contrasts in the efficiency of photolysis and condensation (in the case of NH$_3$).  Because NH$_3$ and PH$_3$ are the most abundant photochemically-active species in the region of Saturn's troposphere above the cloud decks, tropospheric photochemistry revolves around the coupled chemistry of these species \citep[see the reviews of][]{83strobel,05strobel,84atreya,86west,09fouchet}.  Although CH$_4$ is more abundant than either NH$_3$ or PH$_3$, methane is only photolyzed at UV wavelengths shorter than $\sim$145 nm, and these short-wavelength photons are absorbed by other atmospheric constituents well above the tropopause, shielding methane from photolysis in the troposphere.  Moreover, NH$_3$ and PH$_3$ photolysis products do not readily react with CH$_4$, so methane photochemistry is more prominent in the stratosphere than the troposphere. Photolysis and photochemical destruction of H$_2$S may occur within and below the NH$_4$SH cloud deck, but little is known about the ultimate fate of the sulfur photochemical products \citep[see][for details]{76prinn,84lewis_book}.  The same can be said for AsH$_3$ and GeH$_4$ photochemistry \citep[e.g.,][]{85fegley,93nava}.  

\begin{figure*}%
\figurebox{6.0in}{}{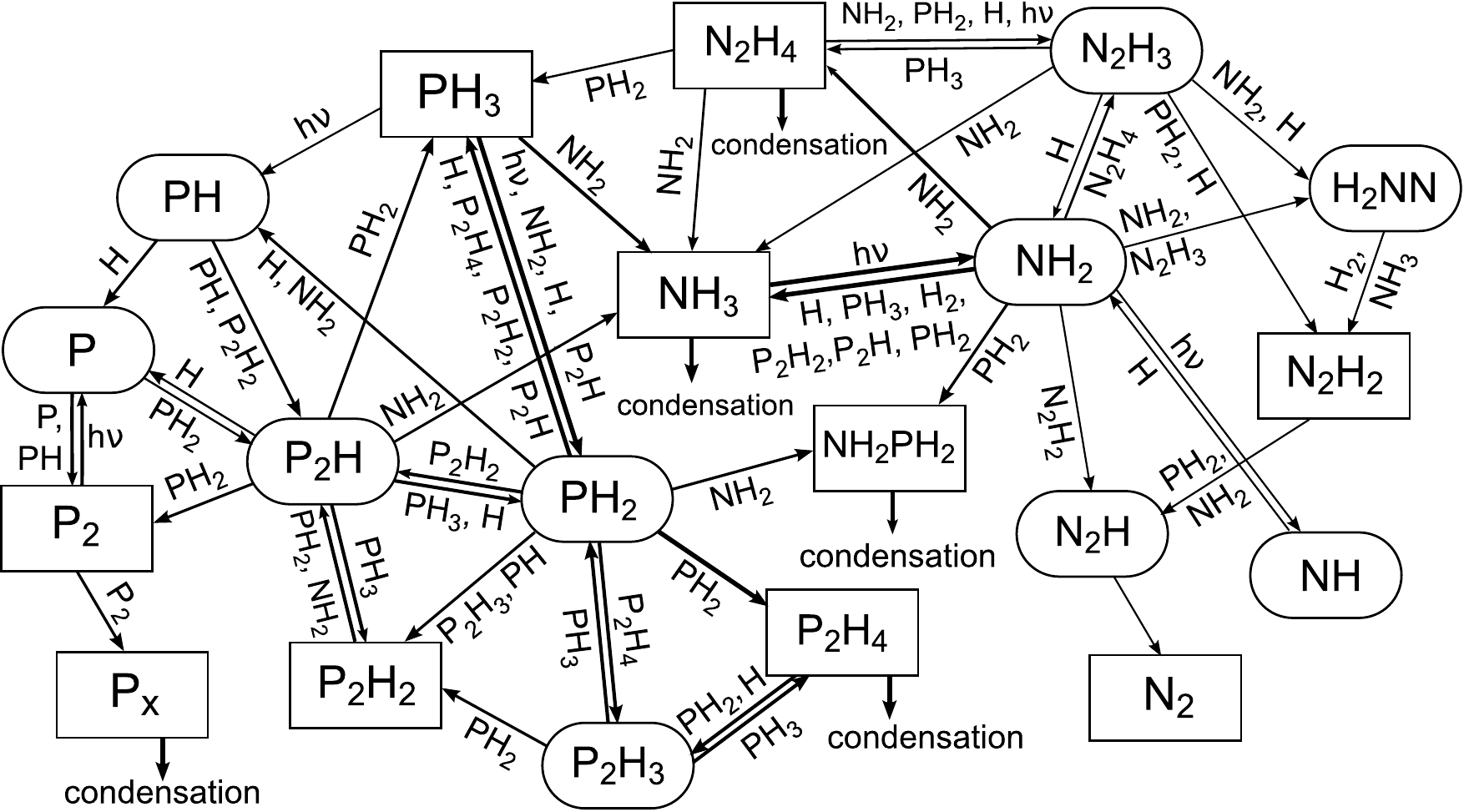}
\caption{Schematic diagram illustrating the important reaction pathways for coupled PH$_3$ and NH$_3$ in Saturn's troposphere \citep[based on][]{09visscher,84kaye}.  The rectangular boxes represent stable molecules, rounded boxes represent radicals or less stable molecules, and h$\nu$ represents ultraviolet photolysis.  The reaction network begins with the photolysis of NH$_3$ and PH$_3$ and the destruction of PH$_3$ from reaction with ammonia photolysis products.  The dominant end products are condensed P$_2$H$_4$, P$_x$ (elemental phosphorus), N$_2$H$_4$, NH$_2$PH$_2$ and gas-phase N$_2$.  Ammonia is also lost via condensation.}
\label{figbubb}
\end{figure*}

Owing to a lack of relevant rate-coefficient information for PH$_3$ photochemistry, in particular, our understanding of tropospheric photochemistry on Saturn (and Jupiter, which is similar) has remained fairly stagnant since the pioneering studies in the late 1970s and 1980s \citep{75strobel,77strobel,77atreya,80atreya,83kaye_hcn,83kaye_ch3nh2,83kaye_ph3,84kaye}. Advances since that time, which mostly revolve around the relevant kinetics and product quantum yields from PH$_3$ and coupled PH$_3$-NH$_3$ and NH$_3$-C$_2$H$_2$ photochemistry, are reviewed by \citet{09fouchet}.  

Ammonia is photolyzed by photons with wavelengths less than $\sim$230 nm; the dominant photolysis products are NH$_2$ and H.  The NH$_2$ reacts effectively with H$_2$ to reform NH$_3$, releasing another hydrogen atom in the process.  Although NH$_3$ condenses in the upper troposphere, its vapor pressure at the tropopause is large enough that NH$_3$ photolysis is an important source of atomic hydrogen in the upper troposphere and lower stratosphere in the models (see Fig.~\ref{figtropchem}), affecting the photochemistry of both local PH$_3$ and the complex hydrocarbons that are flowing down from their production regions at higher altitudes. The NH$_2$ and H can also react with PH$_3$ to form PH$_2$.  The reaction of NH$_2$ with PH$_3$ helps recycle the ammonia.  Two PH$_2$ radicals can recombine to form P$_2$H$_4$, which condenses in its production region, preventing the effective recycling of PH$_3$.  Phosphine itself does not condense under tropospheric conditions on Saturn, but direct photolysis (to produce largely PH$_2$ + H) and reaction of PH$_3$ with NH$_2$ and H serve to drastically decrease the PH$_3$ abundance above the ammonia clouds (see Fig.~\ref{figtropchem}).  

Fig.~\ref{figbubb} shows the details of this coupled NH$_3$-PH$_3$ photochemistry, as described by \citet{09visscher}.  The dominant product is condensed P$_2$H$_4$, which becomes a major component of the upper-tropospheric haze.  Photochemistry of P$_2$H$_4$ can lead to the production of condensed elemental phosphorus as an additional major photochemical product, but the dominant pathways are more speculative \citep{09visscher}.  Other important but less abundant photoproducts include condensed NH$_2$PH$_2$, condensed N$_2$H$_4$, and gas-phase N$_2$ \citep[e.g.,][]{84kaye}.  The N$_2$ shown in Fig.~\ref{figtropchem} is derived from thermochemical quenching and transport from the deep troposphere, with photochemical production contributing only a tiny fraction in comparison with this deep source.  

NH$_3$ photochemistry plays only a minor role in shaping its vertical profile at and below the cloud decks.  Photochemistry is responsible for removing NH$_3$ above the clouds (with N$_2$H$_4$ being the dominant product), but dynamics and condensation control the profile within the cloud region.  Hemispheric contrasts are more strongly influenced by the thermal profile, condensation, regional dynamics (e.g., equatorial upwelling) and aerosol-microphysical processes.  These competing effects have not yet been disentangled to explain the suggested weak enhancement of ammonia in Saturn's southern summer hemisphere in Section \ref{nh3}.  Furthermore, it is not understood whether the NH$_3$ distribution is responding to the seasonal insolation shifts (i.e., following the thermal and aerosol changes), or remaining static with time.  We might expect that northern hemisphere warming will permit the release of more NH$_3$ into the vapor phase during northern spring, helping to reverse the asymmetry observed during southern summer.

The PH$_3$ vertical profile, on the other hand, is strongly sensitive to chemistry, as well as to vertical transport and to the opacity of upper-tropospheric hazes and clouds, which help shield the PH$_3$ from photolysis and other photochemical losses.  The model PH$_3$ vertical profiles drop much more sharply with altitude in the upper troposphere than the profiles derived from CIRS and VIMS retrievals \citep[e.g.,][]{07fletcher_ph3,09fletcher_ph3,11fletcher_vims}, but these differences are likely an artifact of the assumptions in the data retrievals rather than a true model-data mismatch.  Indeed, Herschel/SPIRE observations \citep{12fletcher_spire} suggested that PH$_3$ is not present in the lower stratosphere at the 10-20 mbar level because it is so chemically fragile.  The observed hemispheric asymmetry in the upper-tropospheric PH$_3$ abundance \citep[e.g.,][]{09fletcher_ph3,11fletcher_vims} may be related to higher haze opacities of the UV-shielding aerosols produced by photochemistry in the summer hemisphere, although the global circulation system (i.e., an inter-hemispheric transport from the autumn to the spring hemisphere, see Section \ref{dynamics}) may also play a role.  Seasonal tropospheric photochemistry has not yet been investigated theoretically.  Summertime insolation will promote the production of aerosols like P$_2$H$_4$, which will help shield the PH$_3$ and NH$_3$ from photolysis, allowing the PH$_3$ molecules to be carried to higher altitudes during southern summer and autumn, qualitatively consistent with Cassini's observations.



Unsaturated hydrocarbons like C$_2$H$_2$ are not particularly abundant in the region in which PH$_3$ and NH$_3$ photochemistry is active (see Fig.~\ref{figtropchem}), limiting the effectiveness of the coupled photochemistry of NH$_3$ and PH$_3$ with C$_2$H$_x$ and C$_3$H$_x$ species \citep[e.g.,][]{10moses}, which would otherwise be expected to produce HCN, acetonitrile, methylamine, ethylamine, acetaldazine, acetaldehyde hydrazone, vinylphosphine, ethylphosphine, and a whole suite of other interesting organo-nitrogen and organo-phosphorus species \cite{83kaye_hcn,83kaye_ch3nh2,88ferris,95guillemin,96keane,10moses}. The upper-tropospheric C$_2$H$_2$ abundance is not notably enhanced in latitude regions known to host thunderstorm activity \citep{12hurley}, suggesting that lightning chemistry does not play much of a role in enhancing upper-tropospheric hydrocarbons beyond their photochemically produced abundances.  Based on the chemical mechanism of \citet{10moses}, hydrogen cyanide would be the dominant product of the coupled hydrocarbon-NH$_3$ photochemistry on Saturn, but these models suggest that photochemically produced HCN is not abundant enough to be observable on Saturn (see Fig.~\ref{figtropchem}), and despite a tentative detection of HCN by \citet{96weisstein}, Herschel analyses have provided upper limits an order of magnitude smaller \citep[mole fractions less than $1.6\times10^{-11}$ if the species is well-mixed,][]{12fletcher_spire}. Small amounts of all of the aforementioned organic compounds are produced, however, and will condense, which could potentially contribute to the cloud chromophores on Saturn \citep[see also][]{12carlson}.  Photo-processing of the condensed P$_2$H$_4$, elemental phosphorus, or NH$_4$SH is also a potential source of the yellowish colors on Saturn.  The fact that the yellow-brown coloring is apparently absent in the Cassini images of the high-latitude winter hemisphere of Saturn soon after it emerges back into sunlight --- thought to result from a clearing, thinning, or reduction in size or depth of the aerosols in the winter hemisphere \citep[e.g.,][]{07fletcher_temp,09west} --- further points to a photochemical product as the source of the chromophore (see Section \ref{clouds}).


\subsubsection{Para-hydrogen}
\label{sec_parah2}

\adjustfigure{100pt}

\begin{figure*}%
\figurebox{6.4in}{}{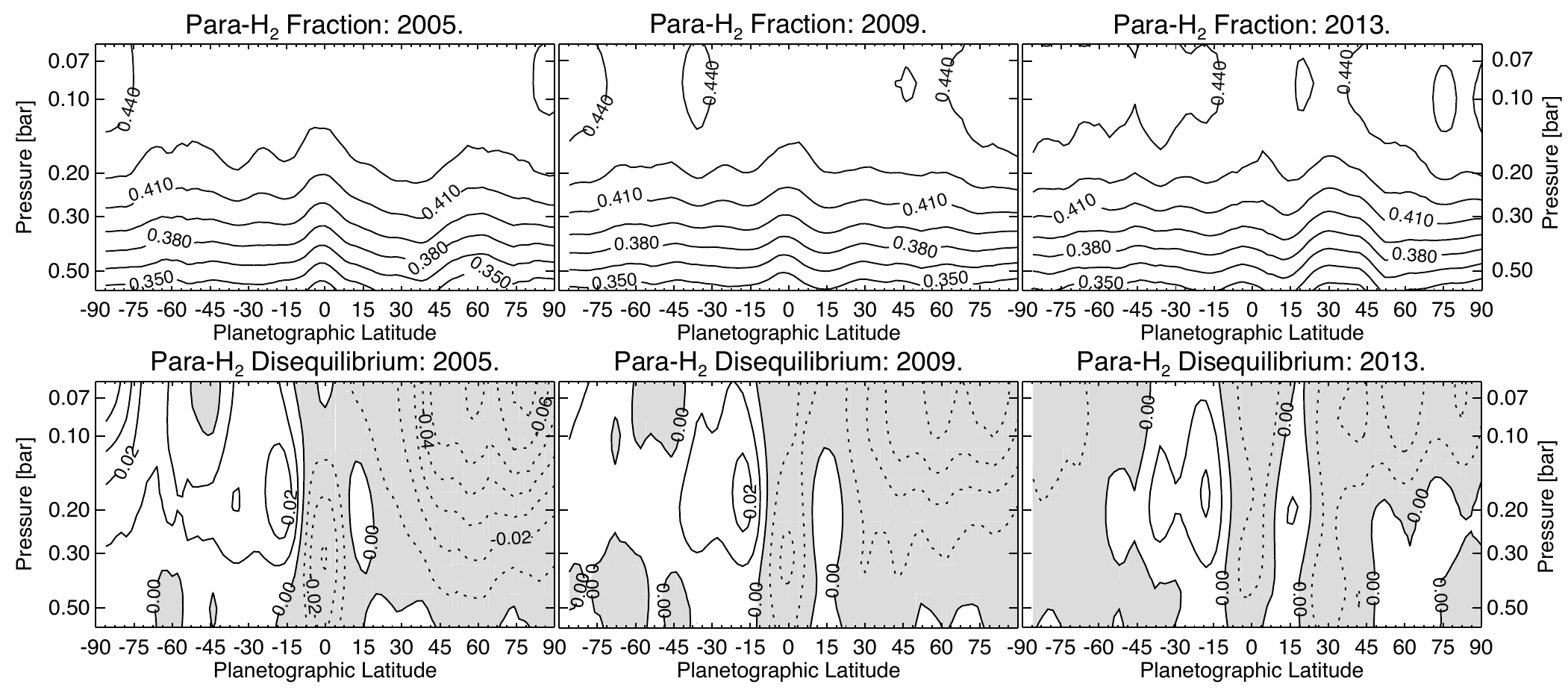}
\caption{Seasonal changes in the zonal mean para-$H_2$ fraction measured by Cassini/CIRS using the collision-induced H$_2$-H$_2$ continuum \citep{07fletcher_temp, 10fletcher_seasons, 15fletcher_fp1}.  The upper panels show the subtle changes in $f_p$ between 2005 and 2013, including the rising $f_p$ at northern high latitudes during spring.  The lower panels compares the degree of disequilibrium between 2005 and 2013 (contours have a 1\% spacing), showing the reversal of the strong asymmetries found in southern summer.  Super-equilibrium conditions ($f_p>f_{eqm}$) are shown as solid lines; sub-equilibrium conditions ($f_p<f_{eqm}$) are shown as dotted lines and light grey shading.}
\label{parah2}
\end{figure*}

Our final tropospheric species shapes the underlying H$_2$-H$_2$ and H$_2$-He collision-induced continuum throughout the infrared and sub-millimeter, in addition to readily identifiable quadrupole line features near 0.62-0.64 $\mu$m and collision-induced dipole features near 0.826 $\mu$m and 2.1-2.2 $\mu$m.  Despite the lack of a permanent dipole, collisions between these molecules create instantaneous dipoles that shape the continuum, and the ratio of ortho-H$_2$ (corresponding to the odd rotational spin state of H$_2$ with parallel spins) and para-H$_2$ (the even spin state of H$_2$ with anti-parallel spins) can be deduced from the relative sizes of the broad S(0) (due to para-H$_2$) and S(1) (due to ortho-H$_2$) absorptions near 354 and 587 cm$^{-1}$, respectively.  In ``normal'' hydrogen, at the high temperatures of Saturn's lower troposphere, this ratio should be 3:1 in equilibrium (a para-H$_2$ fraction $f_p$ of 25\%) \citep[e.g.,][]{82massie}.  At the colder temperatures of the upper troposphere the equilibrium para-H$_2$ fraction ($f_{eqm}$) increases beyond 25\% as para-H$_2$ has the lowest rotational energy state, and we might expect the tropopause para-H$_2$ fraction to be in the 45-50\% range, falling as we rise further into the warm stratosphere.   However, parcels of air displaced upwards will retain their initial low para-H$_2$ fraction if the vertical mixing is faster than the chemical equilibration timescale, leading to sub-equilibrium conditions ($f_p<f_{eqm}$) \citep{98conrath, 07fletcher_temp}.  Conversely, downward displacement of cool air with high values of $f_p$ can lead to super-equilibrium conditions ($f_p>f_{eqm}$).

The spatial distribution of the para-H$_2$ fraction, and its deviation from equilibrium, can therefore be used to trace both vertical mixing and the efficiency of chemical equilibration, both of which may vary with Saturn's seasons.  On Jupiter, where seasonal influences are expected to be negligible, inversions of Voyager/IRIS spectra by \citet{98conrath} revealed a distribution of para-H$_2$ that was largely symmetric about the equator, with evidence for sub-equilibrium conditions at the equator and super-equilibrium at higher latitudes.  Voyager observations of Saturn in early northern spring ($L_s=8.6-18.2^\circ$) revealed local sub-equilibrium conditions near $60^\circ$S, and an asymmetry with a higher $f_p$ in the spring hemisphere \citep[super-equilibrium conditions from 70$^\circ$N to 30$^\circ$S,][]{98conrath}.   These data, most sensitive to the para-H$_2$ fraction between 100-400 mbar (i.e., below the tropopause but above the radiative convective boundary), were interpreted as atmospheric subsidence in the spring hemisphere.

Conversely, early measurements by Cassini during southern summer ($L_s=297-316^\circ$) revealed super-equilibrium over much of the southern summer hemisphere and sub-equilibrium in the winter hemisphere poleward of $30^\circ$N \citep{07fletcher_temp}.  Much of this asymmetry was due to the gradient in $f_{eqm}$ related to the seasonal temperature contrast, whereas $f_p$ itself was found to be largely symmetric about the equator (e.g., left panels of Fig. \ref{parah2}).  \citet{10fletcher_seasons, 15fletcher_fp1} demonstrated that the super-equilibrium region was moving slowly northwards throughout the ten-year span of Cassini observations, both due to an increased $f_p$ and warmer temperatures in the northern spring hemisphere (i.e., smaller $f_{eqm}$).  The para-H$_2$ fraction in the southern hemisphere changed very little between 2004 and 2014, although the cooling temperatures as southern winter approached caused an increase in the expected $f_{eqm}$, thereby decreasing the discrepancy between $f_p$ and equilibrium expectations.  In summary, the stark asymmetry in $f_p-f_{eqm}$ observed during southern summer has reduced significantly (see Fig. \ref{parah2}), such that the southern hemisphere is now close to equilibrium, whereas the wintertime sub-equilibrium conditions still prevail at the highest northern latitudes.  As a result, the zonal mean $f_p$ measured by Cassini in 2014 now qualitatively resembles that determined by Voyager in the last northern spring \citep{15fletcher_fp1}, although quantitative differences remain to be understood, particularly at Saturn's equator.
\adjustfigure{100pt}

Given that the timescales for para-H$_2$ equilibration range from decades to centuries in the troposphere \citep[e.g.,][]{98conrath,03fouchet}, even in the presence of aerosols whose surfaces could provide paramagnetic sites to catalyze the efficient conversion of uplifted para-H$_2$ back to ortho-H$_2$, the discrepancy from a seasonally-dependent equilibrium ($f_p-f_{eqm}$) may not be a good measure of tropospheric circulation.  Put another way, the theoretical $f_{eqm}$ varies instantaneously with temperatures on seasonal timescales, whereas $f_p$ undergoes much more subtle changes with time (showing the largest changes in the northern spring hemisphere).  Furthermore, \citet{07fletcher_temp} speculated that the intricate connection between aerosol catalysts and the rate of para-H$_2$ conversion means that equilibrium conditions would be more easily attained when the atmosphere is hazy (e.g., summertime) than when it is aerosol-free (wintertime), and that this was responsible for the asymmetry observed in $f_p-f_{eqm}$ during southern summer.  In either case, it is the tropospheric temperature and/or aerosol availability that determines the magnitude of the disequilibrium, rather than large-scale overturning. Only in regions of strong localized dynamics is $f_p$ significantly altered from equilibrium conditions (e.g., air rising at the equator, storm eruptions, and air sinking at high northern latitudes during spring).  For example, the equatorial minimum in $f_p$ (sub-equillibrium conditions) has persisted throughout the Cassini mission, consistent with the evidence of powerful equatorial upwelling in the temperature, aerosol, ammonia and phosphine fields \citep{07fletcher_temp, 10fletcher_seasons}.   The connection between the para-H$_2$ fraction and the tropospheric temperatures, aerosols and circulation remains a topic of active exploration.


\subsection{Stratospheric composition}
\label{strat_chem}

In this section, we review spatially-resolved observations and modeling of Saturn's stratospheric hydrocarbons, and their variations with time, to reveal the photochemical and dynamical processes shaping the middle atmosphere.  Stratospheric photochemistry on Saturn is dominated by Lyman-alpha photolysis of methane at high altitudes, producing a multitude of hydrocarbons, many of which have been observed.  Ethane (C$_2$H$_6$) was the first photochemical product of methane to be detected in Saturn's stratosphere from its $\nu_9$ emission band at 12 $\mu$m by \citet{74gillett}, confirmed shortly after by \citet{75tokunaga}.  Acetylene (C$_2$H$_2$) was detected a few years later by \citet{79moos} in the UV using data from the International Ultraviolet Explorer (IUE). A column abundance of ethane was obtained using spectra at 3 $\mu$m by \citet{81Bjoraker}.  The first stratospheric mixing ratios of ethane and acetylene were derived from Voyager/IRIS spectra in the thermal infrared \citep{84courtin, 05sada}. These measurements confirmed C$_2$H$_6$ and C$_2$H$_2$ as the dominant photochemical products.  Constraints from the Infrared Space Observatory (ISO) measurements \citep{97degraauw} were also used by \citet{00moses} to demonstrate that acetylene increases with altitude, as expected for a chemical produced in the upper stratosphere and transported downward by eddy diffusion \citep[see also][]{06prange}.  New hydrocarbon species were identified in the 1990s due to the increasing sensitivity of mid-infrared detectors, leading to the determination of the disk-average integrated column abundance of methlyacetylene (CH$_3$C$_2$H), diacetelene (C$_4$H$_2$) \citep{97degraauw, 00moses}, benzene (C$_6$H$_6$) \citep{01bezard} and the methyl radical \citep{98bezard}, all from ISO. Using ground-based high-resolution spectroscopic data from the NASA/Infrared Telescope Facility (IRTF), ethylene (C$_2$H$_4$) \citep{01bezard_dps} and propane (C$_3$H$_8$) \citep{06greathouse} were also detected.   

Saturn hydrocarbon photochemistry has been recently reviewed by \citet{09fouchet}, and there have been few significant theoretical advances since the time of that review.   A reduced hydrocarbon reaction mechanism suitable for computationally expensive 3D models has been described in \citet{11dobrijevic}, and recent improvements in photochemical models of Titan and the other giant planets are often directly applicable to Saturn \citep[see][and references therein]{11lavvas,11gans,11bell,12vuitton,12westlake,12mandt,12plessis,12moreno,13hebrard,14lara,14dobrijevic,14krasnopolsky,14orton_chem, 15loisin}. Our knowledge of relevant reaction rate coefficients and overall reaction pathways and products is also continually improving due to new laboratory investigations or theoretical calculations (with references too numerous to list). A recent theoretical advance is the new 2D seasonal photochemical modeling of \citet{15hue}, which takes into account the expected seasonal variation in temperatures.  Full details of methane photochemistry on Saturn and the other giant planets can be found in \citet{84atreya}, \citet{83strobel,05strobel}, \citet{96gladstone}, \citet{99yung}, and \citet{00moses,05moses_jup}.

Methane photolysis leads to the production of CH, CH$_2$ (both the ground-state triplet and the excited singlet) and CH$_3$. The CH radicals tend to favor unsaturated hydrocarbon production, whereas the CH$_3$ radicals tend to recycle the methane or lead to the production of saturated hydrocarbons. The peak hydrocarbon production region at 10$^{-3}$--10$^{-4}$ mbar occurs just below the methane homopause where the Lyman alpha photons are absorbed by CH$_4$, but there is a secondary peak in the 0.1-10 mbar region due to photosensitized destruction of CH$_4$ resulting from photolysis of C$_2$H$_2$ and other hydrocarbon photochemical products.  This secondary production region can particularly affect the relative abundances of the hydrocarbons at the pressures at which the infrared observations are sensitive.  

\begin{figure*}%
\figurebox{7.0in}{}{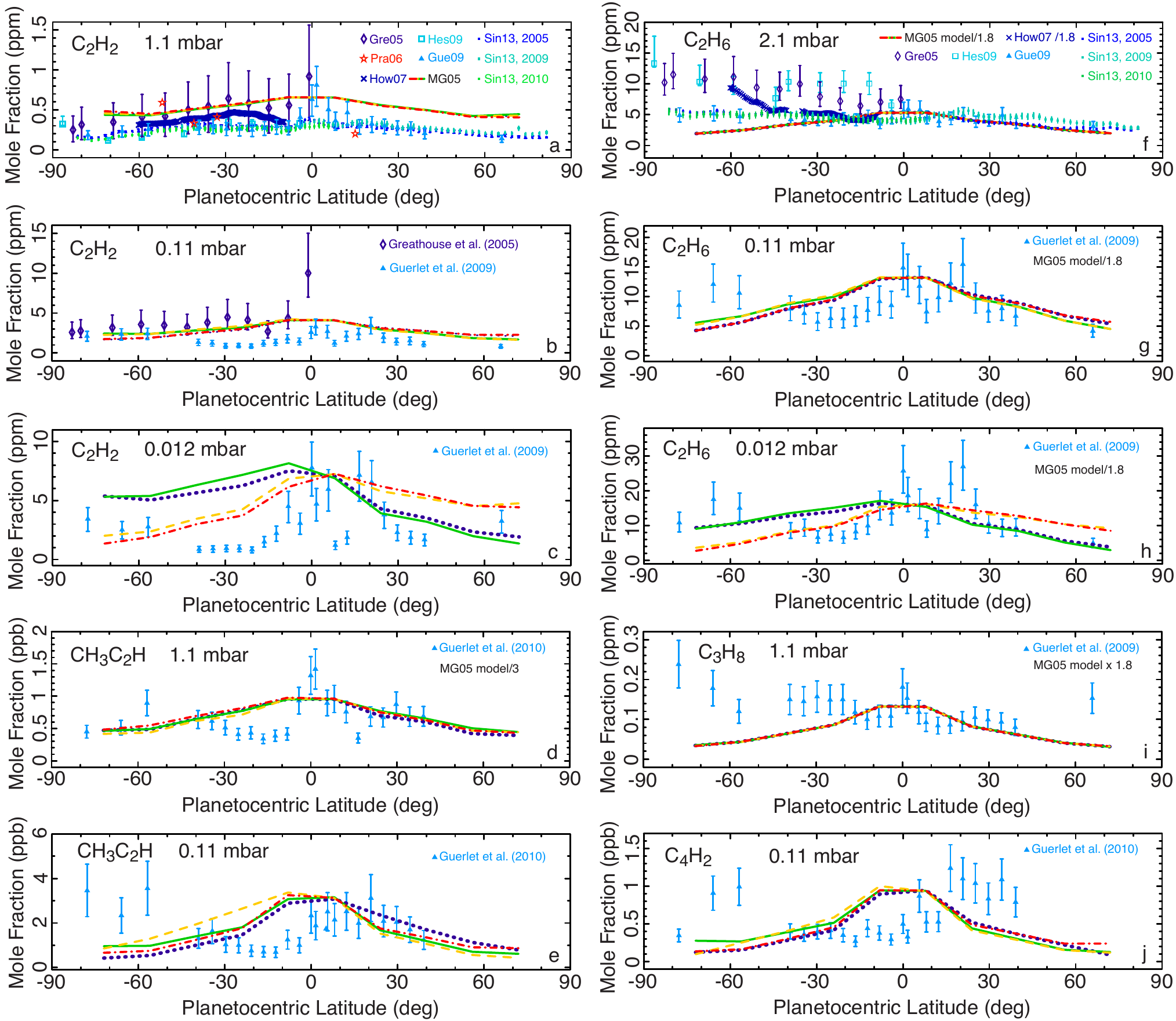}
\caption{Comparing zonal mean hydrocarbon distributions measured for five species (ethane, acetylene, propane, methylacetylene and diacetylene) with the photochemical predictions of \citet{05moses_sat}.  Observers have been abbreviated as follows:  Gre05 \citep{05greathouse}; Pra06 \citep{06prange}; How07 \citep{07howett}; Hes09 \citep{09hesman}; Gue09 \citep{09guerlet}; and Sin13 \citep{13sinclair}.  The model output (labelled MG05) is shown for the solstices and equinoxes to show the predicted magnitude of seasonal variability, particularly for $p<0.1$ mbar:  $L_s=270^\circ$ is dotted (purple), $L_s=0^\circ$ is solid (green), $L_s=90^\circ$ is dashed (orange) and $L_s=180^\circ$ is dot-dashed (red).  The colors of the individual measurements are designed to represent the seasonal timing of the observations.  The model abundances of C$_2$H$_6$, CH$_3$C$_2$H and C$_3$H$_8$ of \citet{05moses_sat} have been approximately scaled to match the data, as described in the figure legends.  Note that C$_3$H$_8$ exhibits a similar meridional trend to C$_2$H$_6$, whereas all the other species tend to track C$_2$H$_2$.}
\label{cxhy}
\end{figure*}

As we shall see, the hydrocarbon abundances vary with both altitude and latitude.  One obvious cause of this variation is seasonal change: the mean daily solar insolation, atmospheric temperatures, and perhaps circulation patterns on Saturn change with season, inducing a response in the vertical and meridional distributions of chemical species. Although three-dimensional dynamical models that include chemistry have not yet been developed for Saturn or the other giant planets, \citet{05moses_sat} have investigated seasonal stratospheric chemistry on Saturn with a 1D time-variable photochemical model, exploring how the molecular abundances change solely due to seasonally-varying solar insolation.   The results of this model are compared to the measured distributions of hydrocarbons in Fig. \ref{cxhy} at a variety of different pressure levels, showing the expected seasonal variability based on photochemistry alone.  This figure will be referred to throughout this section.  The Moses and Greenhouse model includes the effects of ring shadowing and solar-cycle variations, but neglects horizontal transport \citep[e.g.,][]{07moses} or time-variable temperatures and vertical winds.  The hemispheric asymmetries in the hydrocarbon abundances are more pronounced at higher altitudes where vertical diffusion time scales and chemical lifetimes are short.  Our theoretical discussion of the principle production and loss mechanisms for each species will be guided by the results of this 1D model.  Note, also, the recent 2D (latitude and altitude) study by \citet{15hue} that extends the \citet{05moses_sat} study by considering the effects of seasonally varying temperatures.

\subsubsection{C$_2$ hydrocarbons}

Observations and models both demonstrate that C$_2$H$_6$ is the most abundant hydrocarbon photochemical product in Saturn's stratosphere. Ethane is produced predominantly by CH$_3$ recombination throughout the stratosphere and by sequential addition of H to unsaturated C$_2$H$_x$ hydrocarbons in the lower stratosphere; ethane is lost via photolysis and reaction with C$_2$H produced from C$_2$H$_2$ photolysis. Production of C$_2$H$_6$ exceeds loss at most altitudes in the Saturn models, and the net C$_2$H$_6$ production is balanced by downward flow in to the deeper troposphere, where the ethane will eventually be thermochemically converted back to methane. The chemical loss time scale for C$_2$H$_6$ exceeds a Saturn year at most altitudes below its peak production region, making C$_2$H$_6$ particularly stable and thus a good tracer for local dynamics and global circulation.  

Ethylene is produced through the reaction of CH with CH$_4$, reactions of C$_2$H$_3$ with H and H$_2$, and through C$_2$H$_6$ photolysis.  It is lost through H-atom addition to form C$_2$H$_5$ and through photolysis.  \adjustfigure{165pt}Unlike C$_2$H$_6$, ethylene can be photolyzed by UV photons with wavelengths out to $\sim$200 nm, which makes the C$_2$H$_4$ much less stable in the lower stratosphere than ethane.  Its chemical lifetime is shorter than an Earth year at most altitudes, although recycling can help keep it around longer than its loss time scale would indicate. The very low levels of C$_2$H$_4$ under quiescent conditions on Saturn are consistent with photochemical models; its greatly enhanced abundance in the northern hemispheric storm beacon region \citep[][and Chapter 13]{12hesman} is most likely the result of the high dependence of the C$_2$H$_3$ + H$_2$ $\rightarrow$ C$_2$H$_4$ + H reaction on temperature \citep[e.g.,][]{06tautermann,14armstrong,14moses_dps}.

Acetylene is the second-most abundant hydrocarbon photochemical product on Saturn.  Its chemistry is more intricate, complicated, and non-linear, and C$_2$H$_2$ is a particularly important parent molecule for other observed hydrocarbons such as CH$_3$C$_2$H and C$_4$H$_2$.  Acetylene is produced primarily from the photolysis of ethane and ethylene, although ``recycling'' reactions such as C$_2$H + H$_2$ $\rightarrow$ C$_2$H$_2$ + H, H + C$_2$H$_3$ $\rightarrow$ C$_2$H$_2$ + H$_2$, and C$_2$H + CH$_4$ $\rightarrow$ C$_2$H$_2$ + CH$_3$ dominate the total column production rate of C$_2$H$_2$ in the stratosphere.  Acetylene is indeed rapidly recycled throughout the stratosphere, although there is a steady leak out of the ongoing recycling reactions into C$_2$H$_6$, C$_4$H$_2$ and higher-order hydrocarbons. Acetylene has a lifetime intermediate between that of C$_2$H$_6$ and C$_2$H$_4$:  its pure chemical loss time scale is of order of a Saturn season through much of the stratosphere \citep[$\sim$0.03-100 mbar, see][]{05moses_sat}, but recycling reactions give it a much longer ``effective'' lifetime.  


In the absence of atmospheric circulation, photochemically-generated species at microbar pressures (e.g., Fig. \ref{cxhy}) tend to have abundances that follow local insolation conditions, resulting in higher abundances in the summer hemisphere compared to the winter hemisphere \citep{05moses_sat}.  At lower altitudes, however, vertical diffusion time scales and chemical lifetimes are much longer, and the meridional distributions track the yearly average of the mean daily solar insolation rather than the instantaneous solar fluxes. Because the yearly average insolation is largest at the equator and decreases toward the poles, the photochemical models predict that most of the hydrocarbon photochemical products should have a maximum abundance at the equator, decreasing smoothly toward both poles at pressures of 1 mbar and greater.  These predictions can be tested by comparison to latitudinally-resolved hydrocarbon measurements.


\textbf{Ethane and acetylene in southern summer}

\citet{05greathouse} were the first to present latitudinally-resolved distributions of the principle methane photolysis products, ethane and acetylene, in Saturn's southern (summer) hemisphere (Fig. \ref{cxhy}a). Using IRTF data acquired in 2002, they showed that acetylene decreased from equator to pole at the 1-mbar pressure level (in agreement with photochemical model predictions), whereas the opposite trend was marginally observed for ethane at 2 mbar.  These results suggested that dynamical redistribution is effective in Saturn's stratosphere and operates on timescales less than ethane's chemical lifetime of $\sim$700 years and longer than acetylene's lifetime of $\sim$100 years \citep{05moses_sat}. 

Since the work of \citet{05greathouse}, the distributions of ethane and acetylene have been routinely measured by the Composite Infrared Spectrometer (CIRS) onboard Cassini, using a combination of nadir and limb sounding.   Nadir and limb observations provide complementary information. On the one hand, CIRS nadir data provide an excellent spatio-temporal coverage of Saturn's atmosphere and are mainly sensitive to the abundance of ethane and acetylene, with a peak sensitivity near the $\sim2$-mbar pressure level. Under particular conditions, such as the high temperatures encountered within the storm beacon region (see Chapter 13), other trace species can also be measured \citep[such as ethylene, which has been measured by][]{12hesman}. On the other hand, CIRS data acquired in limb-viewing geometry allow the retrieval of the vertical abundance profile of ethane and acetylene over a greater pressure range (typically 5 mbar -- 5 $\mu$bar), but with a much sparser meridional and temporal coverage. 

Five Cassini studies have reported on the meridional distribution of ethane and acetylene during southern summertime based on Cassini/CIRS measurements (Fig. \ref{cxhy}a): three based on nadir data \citep{07howett, 09hesman, 13sinclair} and two based on limb observations \citep{09guerlet,15sylvestre}. Although these studies do not necessarily agree quantitatively, they report similar trends. \citet{07howett} measured the southern hemisphere hydrocarbon distributions in 2004 between $10-70^\circ$S, showing a decrease of acetylene poleward of $30^\circ$S. They also confirmed that the ethane abundance is relatively uniform with latitude equatorward of $50^\circ$S, but suggested a rather large increase poleward of $50^\circ$S.  This increase has not been reproduced by subsequent investigators, and could be ascribed to a bias with emission angle (as the higher latitude data were acquired at higher emission angles) linked to spectroscopic errors.    Furthermore, all follow-up studies based on CIRS measurements rely on an updated ethane linelist by \citet{07vander}, which has line intensities higher by approximately 30\% than the values used by \citet{07howett}.  \citet{09hesman} then extended the CIRS retrievals towards the south pole in 2005, and combined them with low-latitude ground-based observations.  They confirmed acetylene's poleward decrease but found a sharp rise in C$_2$H$_2$ right at the summer pole (87S), within the stratospheric vortex identified by \citet{05orton} and \citet{08fletcher_poles}.  This sharp increase has also been observed for ethane \citep{15fletcher_poles}, and suggests enhancement by strong subsidence within the summer polar vortex. Using limb observations acquired in 2005-2006 covering both hemispheres ($80^\circ$S--$45^\circ$N), \citet{09guerlet} confirmed that ethane and acetylene follow different meridional trends in the 1--5 mbar pressure range: while acetylene generally decreases from the equator towards both poles, the ethane distribution is much more homogeneous with latitude.  These results are also in agreement with the study of \citet{13sinclair} from CIRS nadir observations covering both hemispheres.

%

Figs. \ref{cxhy}a and  \ref{cxhy}f compare the meridional distribution of ethane and acetylene at the 1-2 mbar pressure level during southern summer, as measured by various authors, and as predicted by the photochemical model of \citet{05moses_sat}. The chemistry of C$_2$H$_2$ and C$_2$H$_6$ is highly linked, such that the C$_2$H$_2$ meridional distribution tends to track the C$_2$H$_6$ distribution in photochemical models \citep[e.g.,][]{05moses_sat,07moses,15hue}. The notable differences in the meridional profiles of C$_2$H$_2$ and C$_2$H$_6$ on Saturn is therefore a surprise and suggests the influence of meridional transport on long timescales, as already proposed by \citet{05greathouse}.  \citet{09guerlet} also report a sharp and narrow equatorial maximum in the C$_2$H$_2$ distribution at 1 mbar (and a more moderate local maximum of C$_2$H$_6$), much higher than predicted by the seasonal photochemical model. The authors interpret this strong maximum as the signature of a local subsidence associated with the equatorial oscillation. Indeed, given that C$_2$H$_2$ mixing ratio increases with height, a downwelling wind would carry C$_2$H$_2$ and enrich lower altitude regions.

At sub-millibar pressure levels, \citet{09guerlet} took advantage of limb-viewing geometries from Cassini to derive the distribution of acetylene and ethane in the upper stratosphere (p $<$ 0.1 mbar, Fig. \ref{cxhy}b,c,g,h). Contrary to the trends observed in the lower stratosphere, they find that both species follow very similar trends in the upper stratosphere. These distributions of the two hydrocarbons are asymmetric, with volume mixing ratios 2 to 6 times higher at northern mid-latitudes (with a local maximum centered at $25^\circ$N) than at southern mid-latitudes.  This is opposite to what might be expected from hydrocarbon production rates alone, where production rates should have been higher in the summer southern hemisphere than in the winter hemisphere (see Fig. \ref{cxhy}c,h).   \citet{09guerlet} hence suggest that a strong meridional transport from the summer to the winter hemisphere, occurring on seasonal timescales, is responsible for the observed enrichment at $25^\circ$N.  This hypothesis is consistent with the predictions of the global circulation model of \citet{12friedson}, in which stratospheric circulation cells develop, with a descending branch at $25^\circ$N at this season.

\textbf{Temporal trends in ethane and acetylene}

The southern summer observations by Cassini revealed asymmetries in ethane and acetylene that would be expected to shift over time, potentially reversing by northern summer solstice in 2017 ($L_s=90^\circ$).  \citet{13sinclair} were the first to study temporal variability in the distributions of ethane and acetylene using nadir sounding at low spectral resolution (15~cm$^{-1}$) between 2005 and 2010 ($L_s=308-15^\circ$).  They confirmed (i) the absence of a sharp south-poleward rise in ethane as first reported by \citet{07howett} from early CIRS nadir data ($L_s\approx295^\circ$); and (ii) the equatorial peak abundance of acetylene observed in 2005 by \citet{09guerlet}, although this maximum appears more muted in the nadir data (see Fig. \ref{cxhy}a).  Differences between limb and nadir results in the equatorial region could be ascribed to the temporal evolution of the equatorial oscillation, which strongly perturbs the thermal (and chemical) structure of the stratosphere on small vertical scales not fully resolved with nadir, low spectral resolution data.

\begin{figure}%
\figurebox{3.6in}{}{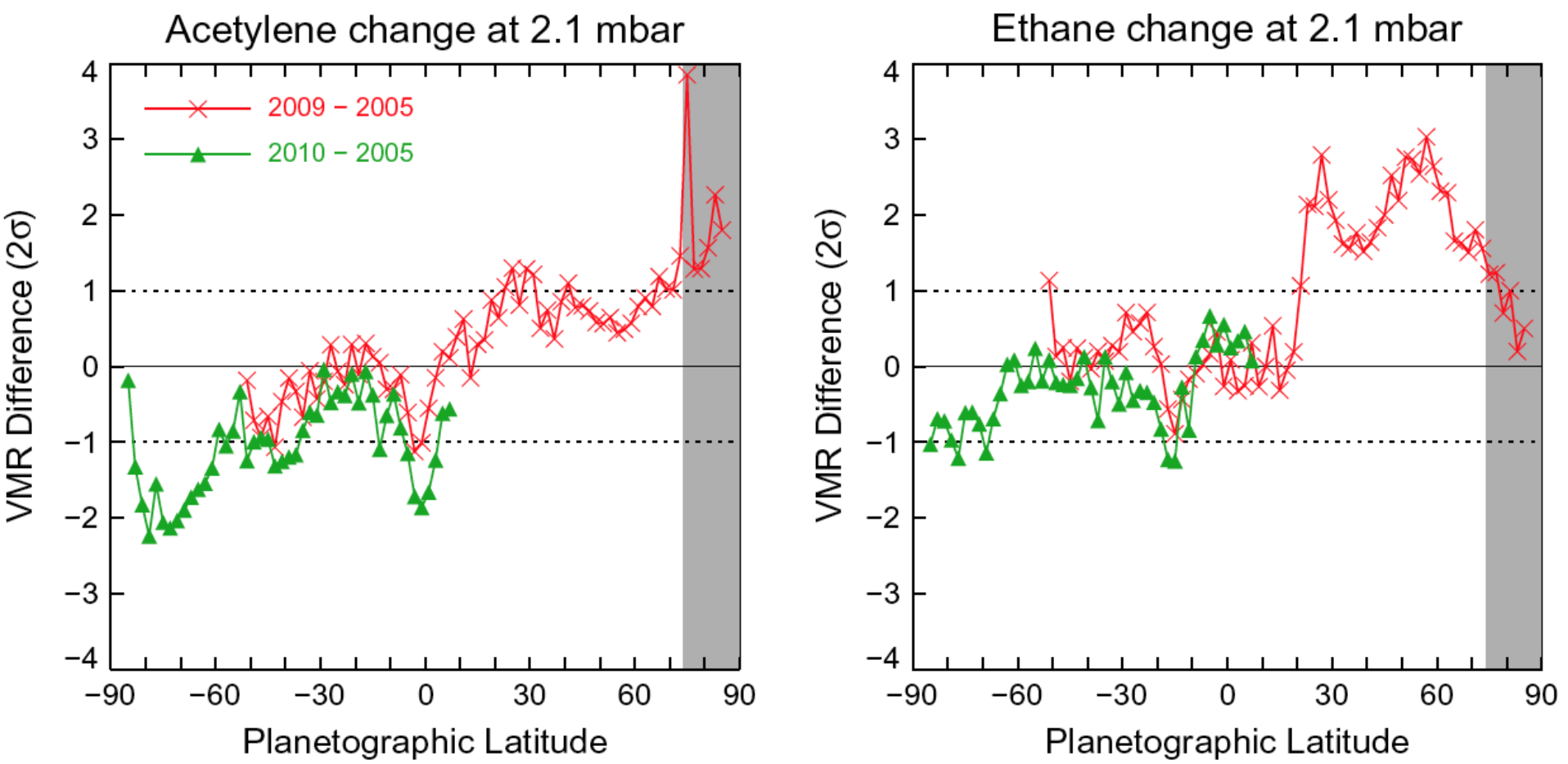}
\caption{Differences in zonal mean retrievals of ethane and acetylene at 2 mbar between 2005 and 2009/2010 from Cassini/CIRS, as presented by \citet{13sinclair}.  Trends are shown in units of the uncertainty ($\sigma$) to reveal where the trends are statistically significant.  In general, the results show a decline in abundance over the southern hemisphere and an increase over the northern hemisphere, suggestive of large-scale inter hemispheric circulation with subsidence in northern winter/spring.  Similar conclusions have been reached using temporal trends in C$_2$H$_2$ and C$_2$H$_6$ distributions at higher latitudes \citep{15fletcher_poles}.  }
\label{cxhy_change}
\end{figure}

Away from the equator, the Cassini/CIRS observations are beginning to reveal seasonal shifts in the zonal-mean ethane and acetylene distributions, as shown in Fig. \ref{cxhy_change}.  At mid-latitudes, \citet{13sinclair} report significant enrichments in ethane (and to a lesser extent, in acetylene) at the 2-mbar pressure level in the region $20-65^\circ$N and depletions at $15^\circ$S, which they interpret as extended subsidence occurring in the northern hemisphere between 2005 and 2010, and upwelling at $15^\circ$S.  \citet{13sinclair} also showed that the equatorial maximum of C$_2$H$_2$ decreases over time (while the ethane concentration remains fairly constant), possibly due to the evolution of the equatorial oscillation.   At higher latitudes, \citet{15fletcher_poles} use ten years of CIRS nadir data to demonstrate that the summertime south polar maximum in both C$_2$H$_2$ and C$_2$H$_6$ has been declining throughout the timespan of the observations ($L_s=293-54^\circ$) at 1-2 mbar, mirrored by enhancements in both species in the north polar spring (poleward of $75^\circ$N).  These changes in the millibar region cannot be explained solely in terms of photochemistry, so that the authors invoke stratospheric circulation from south to north (broadly speaking, upwelling in the autumnal hemisphere and subsidence in the spring hemisphere) to explain the observed changes.  

In many of these studies \citep{09guerlet, 13sinclair, 15fletcher_poles}, the temporal behavior of C$_2$H$_2$ and C$_2$H$_6$ were found to be different (ethane variations appeared more significant than those of acetylene), despite the highly linked chemistry of the two species, and suggestive of the influence of dynamics.  Acetylene is expected to behave differently from ethane in the presence of vertical winds, for example.  In the $\sim$0.1-10 mbar region, downwelling winds could carry more C$_2$H$_2$ from high altitudes to lower altitudes, but the increased C$_2$H$_2$ at these lower altitudes results in a non-linear chemical loss rate due to the increased photolysis (and thus C$_2$H and H production) in conjunction with the resulting increased rates of the non-recycling loss reactions such as C$_2$H + C$_2$H$_2$ $\rightarrow$ C$_4$H$_2$ + H and H + C$_2$H$_2$ + M $\rightarrow$ C$_2$H$_3$ + M.  Thus, the subsidence will lead to an increased fraction of the C$_2$H$_2$ photolysis products being permanently converted to other hydrocarbons rather than recycling the acetylene, resulting in less of an increase than one would expect based on subsidence alone without chemical coupling.  By the same token, upwelling winds will decrease the C$_2$H$_2$ abundance and loss rate non-linearly, favoring recycling over permanent loss.  Upwelling and downwelling winds can therefore lead to less significant decreases and increases in the C$_2$H$_2$ mixing ratios as compared to C$_2$H$_6$, which might help explain some of the temporal behavior observed by \citet{13sinclair}; however, the slope of the unperturbed mixing-ratio profile also contributes to the observed magnitude of the increase or decrease during the upwelling/downwelling \citep[e.g.,][]{09guerlet}, so the resulting response will be complicated.  Coupled 2D and 3D photochemical-dynamical models will likely be needed to fully explain the temporal and meridional variations of stratospheric hydrocarbons in the Cassini CIRS observations.  No such models have been published to date.


At higher altitudes, seasonal changes in the hydrocarbon distributions have also been studied by \citet{15sylvestre}, who analyzed CIRS limb data acquired in 2010-2012 and compared those results to the 2005-2006 observations by \citet{09guerlet}. Contrary to \citet{13sinclair}, \citet{15sylvestre} find little change in the ethane and acetylene distribution in the millibar region. However, they find that the local enrichment in ethane and acetylene previously observed at $25^\circ$N in 2005--2006 (especially at p$<$0.1 mbar) has since disappeared.  If subsidence at 25$^\circ$N was occurring during the northern winter, then it has stopped by northern spring, qualitatively consistent with the predictions of the seasonally-reversing Hadley circulation proposed by \citet{12friedson}.   Alternatively, the variability at 25$^\circ$N could be coupled to secondary meridional circulations induced by the quasi-periodic equatorial oscillations whose stacked pattern is descending with time.

Despite quantitative differences between spatial distributions and temporal behavior identified by different authors, the general trends for ethane and acetylene are consistent:  (i) C$_2$H$_2$ decreasing from equator to pole whereas C$_2$H$_6$ is largely uniform with latitude; (ii) regional enhancements of both species at the highest latitudes within polar vortices, possibly associated with auroral chemistry and dynamic entrainment; (iii) general spring hemisphere enhancements and autumn hemisphere depletions in the millibar region, suggestive of stratospheric circulation from the southern to the northern hemisphere; and (iv) complex temporal behavior in the tropics associated with vertical propagating waves and a seasonally-reversing Hadley-type circulation.  As ethane and acetylene are also the principal stratospheric coolants, their spatial and temporal behavior must also be incorporated in radiative climate models to better explain the thermal asymmetries observed by Cassini (see Section \ref{temp}).

\subsubsection{C$_3$ and higher hydrocarbons}

In addition to ethane and acetylene, Cassini/CIRS has also revealed hemispheric asymmetries in the higher order hydrocarbons.  Although studies of their temporal evolution are rather limited at the time of writing, we review here the main conclusions from both theory and observations.  

\textbf{Propane}:  Propane is produced in the photochemical models through termolecular reactions, which are more effective at higher pressures. Hence, the production rate for C$_3$H$_8$ peaks near $\sim$3 mbar, where photosensitized CH$_4$ destruction occurs, rather than at higher altitudes where CH$_4$ is destroyed by Lyman alpha photolysis. Propane is lost predominantly through photolysis.  Because it is shielded to some extent by C$_2$H$_6$ and C$_2$H$_2$, propane is relatively stable in the models.  Fig. \ref{cxhy}i indicates that propane is expected to show a similar latitudinal trend to C$_2$H$_2$, with a decline from equator to pole in both hemispheres.  However, the first spatial distribution of propane provided by \citet{06greathouse} found that propane at the 5-mbar level was relatively uniform in the southern hemisphere (based on two measurements at $20^\circ$S and $80^\circ$S).  \citet{09guerlet} derived propane from Cassini/CIRS limb observations between March 2005 and January 2008 ($L_s=300-340^\circ$), with a peak sensitivity near 1-mbar. They pointed out that the meridional behavior of C$_3$H$_8$ closely tracks the instantaneous solar insolation (i.e., more propane in the summer hemisphere), suggesting that the chemical lifetime of C$_3$H$_8$ is shorter than the models indicate.  Another possibility is that the C$_3$H$_8$ abundance is sensitive to temperature through reactions that have a higher degree of temperature dependence than is assumed in the models \citep{11dobrijevic}.  In fact, models tend to underpredict the C$_3$H$_8$ abundance on Saturn \citep{06greathouse,09guerlet,11dobrijevic}. Most recently, Cassini/CIRS limb observations in 2010 \citep{15sylvestre} suggest little to no change since 2005-2006.

\textbf{Methylacetylene  and diacetylene:} \citet{10guerlet} continued to exploit the same Cassini/CIRS limb data ($L_s=300-340^\circ$) to provide the first latitudinally-resolved distributions of methylacetylene (CH$_3$C$_2$H) and diacetylene (C$_4$H$_2$), finding that mid-southern latitudes are depleted in both hydrocarbons compared to mid-northern latitudes during southern summer. The photochemical models do not predict such a hemispheric asymmetry during this season (see Fig. \ref{cxhy}), and \citet{10guerlet} suggest that the behavior is caused by upwelling at mid-southern latitudes and subsidence at mid-northern latitudes.  Methylacetylene and diacetylene are important hydrocarbon photochemical products on Saturn that have much shorter chemical lifetimes than C$_2$H$_6$, C$_2$H$_2$, and C$_3$H$_8$. In the upper atmosphere where CH$_4$ is photolyzed by Lyman $\alpha$ and other short-wavelength radiation, CH insertion into C$_2$H$_6$, C$_2$H$_2$, and C$_2$H$_4$ initiates the production of C$_3$H$_x$ hydrocarbons, which then can be photolyzed or react with hydrogen or other species to eventually produce CH$_3$C$_2$H \citep[see][for details]{00moses,05moses_jup}.  At pressures greater than $\sim$0.01 mbar, CH$_3$C$_2$H has additional sources through photolysis of C$_4$H$_x$ species and reactions such as CH$_3$ + C$_2$H$_3$,  which produce C$_3$H$_x$ species that can again make their way to forming CH$_3$C$_2$H.  In this lower-altitude region, acetylene is an important ``parent'' molecule for CH$_3$C$_2$H.  Methylacetylene is lost through photolysis and H-atom addition. The primary (non-recycling) mechanism for C$_4$H$_2$ production is acetylene photolysis followed by C$_2$H + C$_2$H$_2$ $\rightarrow$ C$_4$H$_2$ + H.   Diacetylene formation therefore depends non-linearly on the acetylene abundance. H-atom addition to form C$_4$H$_3$ is the dominant loss reaction for C$_4$H$_2$, but the resulting C$_4$H$_3$ radical can react with H to recycle the C$_4$H$_2$.  Photolysis is an effective loss process, although it, too, leads to some C$_4$H$_2$ recycling.  Diacetylene is expected to condense in the lower stratosphere to contribute to a substantial fraction of the stratospheric haze burden \citep[e.g.,][]{00mosesO}.  

The meridional distribution of CH$_3$C$_2$H (Fig. \ref{cxhy}d-e)and C$_4$H$_2$ (Fig. \ref{cxhy}j) in the 0.05-1 mbar region appears to grossly track that of C$_2$H$_2$ \citep{10guerlet}, which makes sense  theoretically given the short lifetimes of methylacetylene and diacetylene and given that C$_2$H$_2$ is the key ``parent'' molecule in the middle and lower stratosphere for both these  species.  The C$_4$H$_2$ abundance is typically well reproduced in 1D global-average models when the predicted C$_2$H$_2$ profiles match observations, suggesting that the C$_4$H$_2$ chemistry is well understood, but global-average photochemical models tend to notably overpredict the CH$_3$C$_2$H abundance in the 0.1-1 mbar region \citep[e.g.,][]{00moses,05moses_jup,10guerlet,11dobrijevic}, suggesting that the CH$_3$C$_2$H chemistry is not well described in the models. The meridional variations of CH$_3$C$_2$H and C$_4$H$_2$ appear more extreme than for C$_2$H$_2$, which is likely the result of the non-linear nature of their dependence on C$_2$H$_2$ photochemistry; however, these species also exhibit smaller-scale variations and differences between their respective distributions that are not easily explained with the chemical models.  \citet{11dobrijevic} demonstrate that rate-coefficient uncertainties lead to a large spread in the predicted C$_3$H$_8$, CH$_3$C$_2$H and C$_4$H$_2$ abundances in the photochemical models, which could explain these model-data mismatches.  \citet{10guerlet} suggest that seasonally variable transport is affecting the C$_4$H$_2$ and CH$_3$C$_2$H distributions.  

\textbf{Benzene: } Benzene photochemistry is not well understood under conditions relevant to Saturn. Neutral C$_6$H$_6$ photochemical production and loss mechanisms on the giant planets are discussed by \citet{00moses,05moses_jup} and \citet{05lebonnois}. Recent CIRS measurements of the C$_6$H$_6$ column abundance show that benzene is at least as abundant in the polar stratosphere ($80^\circ$S) compared to equatorial regions, in contradiction with photochemical model predictions including only neutral chemistry \citep{15guerlet}. The latter authors conclude that ion chemistry in the auroral regions plays an important role in the benzene production rates, as is the case on Jupiter \citep{00wong,02friedson,03wong} and for non-auroral ion chemistry on Titan \citep{09vuitton}.  The apparent presence of high-altitude hazes on Saturn observed during VIMS stellar occultations \citep{09bellucci,12kim} and by CIRS \citep{15guerlet} further suggests that Titan-like ion chemistry is contributing to the production of complex organics on Saturn. Benzene, like C$_4$H$_2$, is also expected to condense to form haze particles in Saturn's lower stratosphere.  

In summary, Cassini has determined the spatial distribution of Saturn's stratospheric hydrocarbons (ethane, acetylene, propane, methylacetylene, diacetylene and benzene), discovering asymmetries during summertime conditions that bear little resemblance to the photochemical model predictions in the absence of circulation.  Auroral chemistry is a possible contributor to increased abundances in the high-latitude regions, which should be incorporated into the radiative budget in the polar regions. Coupled 2D and 3D photochemical-dynamical models will likely be needed to fully explain the temporal and meridional variations of stratospheric hydrocarbons in the Cassini CIRS observations, but no such models have been published to date.

\subsubsection{Oxygen species}
\label{oxygen}

Finally, the reducing nature of Saturn's stratospheric composition is perturbed by a steady influx of oxygenated species from external sources, such as micrometeoroid precipitation, cometary impacts or a connection with Saturn's rings and satellites.  These oxygen compounds can generate new photochemical pathways to form unexpected molecules, attenuate UV flux or provide condensation nuclei \citep{00mosesO}.  The history of ground-based observations of oxygen compounds, using rotational lines in the far-IR and sub-millimeter, was reviewed by \citet{09fouchet}.  Since the time of that review, investigators have continued to study disk-integrated stratospheric CO and H$_2$O from ground-based \citep{09cavalie,10cavalie} and space-based \citep[Herschel,][]{11hartogh,12fletcher_spire} observatories, and Cassini/CIRS has demonstrated an absence of latitudinal CO$_2$ trends for $p<10$ mbar \citep{13abbas}.  

However, we discuss oxygen compounds briefly here because, although the chemical and diffusion timescales for CO, H$_2$O and CO$_2$ are too long to generate seasonal asymmetries, the water column abundance itself should change with season due to the altering thermal structure.  Because water condenses relatively high in the stratosphere, the far-IR and sub-mm observations will be very sensitive to the altitude at which the water condenses, which is in turn controlled by the stratospheric temperatures.  If the source is isotropic (interplanetary dust) or favors low latitudes \citep[e.g., from Enceladus,][]{11hartogh}, we would expect to see a larger water column in the summer hemisphere and/or wherever mid-to-low stratospheric temperatures are highest.  If the source is from the ring atmosphere \citep{13odonoghue, 11tseng}, then the source itself is seasonal (greater ring atmosphere and thus greater influx when the ring has its highest opening angle at the solstices).  The external oxygen supply is discussed in Chapter 9.

\section{Clouds and hazes}
\label{clouds}

Cloud and haze particles provide yet another potential avenue to probe important atmospheric processes, with both obvious and subtle ties to the wind, temperature, and chemical tracer fields.  Although observation of clouds and haze is relatively straightforward, retrieval of vertical profiles, particle size, shape and composition and relationships to the other fields is not.  Both types of particles can be classified generically as aerosols.  We usually refer to `clouds' in the context of condensation/sublimation of volatile constituents (NH$_3$, NH$_4$SH and H$_2$O/NH$_3$) in the upper troposphere, while the term `haze' is used in association with the solid phase of photochemical products:  P$_2$H$_4$ is expected to exist in the upper troposphere (although this has yet to be confirmed by observations); hydrocarbons and condensed external water in the stratosphere \citep{00moses}.  The two may combine: stratospheric haze sedimenting from high altitudes may serve as condensation nuclei for condensates deeper down.  Conversely, ice crystals in tropospheric clouds may acquire a photochemically-produced hydrocarbon coating.  The reader is referred to \citet{09west} for a comprehensive review of the literature on clouds and hazes, as a full treatment of aerosol composition and derived properties is beyond the scope of this chapter.  

Common features of the various reflectivity investigations \citep[e.g.,][]{92karkoschka, 93karkoschka, 01stam, 05temma, 05perez-hoyos, 05karkoschka} include (a) a stratospheric haze ($1<p<90$ mbar) of small radius ($r\approx0.1-0.2$ $\mu$m) particles, possible originating from photochemical processes; (b) a tropospheric haze from the tropopause down to the first condensation cloud deck at 1.5-2.0 bar, possibly with aerosol-free gaps in the vertical distribution; and (c) a possible thick NH$_3$ cloud, although signatures of fresh NH$_3$ ice have so far only been detected in Saturn's 2010-11 storm region \citep{13sromovsky}.  Numerous explanations have been presented for the concealment of condensate signatures, including large particle sizes, non-spherical particle shapes, coating of the pure ices by photochemical products sedimenting downwards \citep{05atreya, 08kalogerakis}, or by mixing of the condensate phases to mask the spectral signatures.  The polar stratospheric haze appears distinct from all other latitudes, being optically thicker and darker at UV wavelengths, with strong Rayleigh-like polarisation suggestive of the importance of auroral processes in their formation.  High spatial resolution limb images have revealed distinct haze layers at some latitudes \citep[e.g.,][]{12rages}.  The equatorial zone, between $\pm18^\circ$ latitude, features consistently high clouds with perturbations from major storms \citep{05perez-hoyos}.   The vertical distributions of the tropospheric aerosols have been well-mapped with ground-based and Hubble data, and Cassini is beginning to add to this picture.   Here we focus on the seasonal and some non-seasonal behavior of clouds and haze on the scale of the zonal jets.

\subsection{Pre-Cassini reflectivity studies}

Remote sensing measurements of clouds and haze probe the upper troposphere and stratosphere, and it is in this regime where seasonal effects from atmospheric heating and photochemistry are most likely to have an effect.  CCD images in near-infrared methane bands provide a measure of cloud top altitude and haze/cloud density, and it is not surprising that seasonally varying hemispheric asymmetries are apparent in ground-based and space-based images in methane absorption bands.  Other diagnostics of cloud and haze vertical structure include polarization at phase angles near 90 degrees and ultraviolet reflectivity.  Polarization is sensitive to cloud altitude provided that it is dominated by Rayleigh scattering above a non-polarizing cloud.  This is the case for low and middle latitudes but not so for high latitudes on Saturn and Jupiter at blue wavelengths.  At high latitudes haze particles also provide strong polarization.  Ultraviolet reflectivity is diagnostic of cloud altitude provided that it is also dominated by Rayleigh scattering from gas.  Again, this seems to hold reasonably well at low latitude but not at high latitudes where UV-absorbing particles are abundant in the stratosphere.  The latitudinal behavior of the highly-polarizing, UV-absorbing stratospheric aerosols is most likely a result of auroral energy deposition at high latitudes \citep{91pryor, 91west}; polar aerosols are discussed in Chapter 12.

The record of reflectivity measurements now span more than an entire seasonal cycle on Saturn.  The Pioneer 11 flyby occurred just prior to northern spring equinox ($L_s=354^\circ$); the Voyager 1 and 2 encounters were a little after northern spring equinox ($L_s=8.6-18.2^\circ$, respectively).  Near the northern spring equinox in the early 1980s, methane-band imagery from 1979 \citep{82west}, polarization imagery from Pioneer 11 in 1979, and Voyager 2 photopolarimeter scans in 1981 \citep{82lane} all showed hemispheric contrasts at mid-latitudes consistent with deeper clouds and/or a smaller column density of haze in the southern mid-latitudes during early northern spring.  \citet{84tomasko} summarized those results using simple cloud/haze models showing effective cloud top pressures at $40^\circ$N and $40^\circ$S to be near 300 and 460 mbar, respectively.  Viewing geometry and spatial resolution limited the northern analysis to latitudes less than $50-60^\circ$N, while at southern latitudes the methane-band imagery indicated the effective cloud top pressure rising poleward of $40^\circ$S to 320 mbar at $70^\circ$S.  The observed hemispheric difference was initially attributed to higher static stability and suppressed convection at southern latitudes after an extended period of solar heating during southern summer.

The equinoctial snapshot provided by Pioneer and Voyager was supplemented by ground- and space-based observations prior to Cassini's observations, most notably using the Hubble Space Telescope.  During northern spring and summer, \citet{92karkoschka} studied ground-based images from 1986 to 1989; and \citet{93karkoschka} presented some of the earliest Hubble imaging in 1991 ($L_s=68-130^\circ$).  \citet{01stam} presented ground-based images of Saturn near the northern autumn equinox in 1995 ($L_s=176^\circ$), able to view both northern and southern hemispheres simultaneously.  A seasonal asymmetry was observed, with both the tropospheric and stratospheric optical thicknesses appearing larger over northern mid-latitudes (approaching autumn) than over southern mid-latitudes (approaching spring), interpreted by \citet{01stam} as a thicker tropospheric cloud deck in the autumn hemisphere.    

As the northern hemisphere receded from view and Saturn approached southern summer solstice, images from the Hubble Space Telescope (HST) provided a data set of significant value to the study of seasonal and non-seasonal behavior of Saturn's haze and clouds.  \citet{05perez-hoyos,06perez-hoyos} used HST data to study the equator and southern hemisphere between 1994-2003 ($L_s=158-286^\circ$); and \citet{05karkoschka} presented analyses of 134 images of Saturn taken by the Hubble Space Telescope between 1991 and 2004 ($L_s=130-289^\circ$), and performed a principal-component analysis of many latitudes on Saturn.  Four statistically-meaningful principal components emerged. The first principal variation is a strong mid-latitude variation of the aerosol optical depth in the upper troposphere. This structure shifts with Saturn's seasons, but the structure on small scales of latitude stays constant. This is what is most apparent in a casual comparison of images taken in different seasons. The second principal variation is a variable optical depth of stratospheric aerosols. The optical depth is large at the poles and small at mid- and low latitudes, with a steep gradient in between. This structure remains essentially constant in time. The third principal variation is a variation in the tropospheric aerosol size, which has only shallow gradients with latitude, but large seasonal variations.  Aerosols are largest in the summer and smallest in the winter, broadly consistent with the 1980s-equinox observation of a haze free southern autumn hemisphere. The fourth principal variation is a feature of the tropospheric aerosols with irregular latitudinal structure and fast variability, on the time scale of months.  

\subsection{Cassini's observations of seasonal aerosol changes}

\begin{figure}%
\figurebox{3.2in}{}{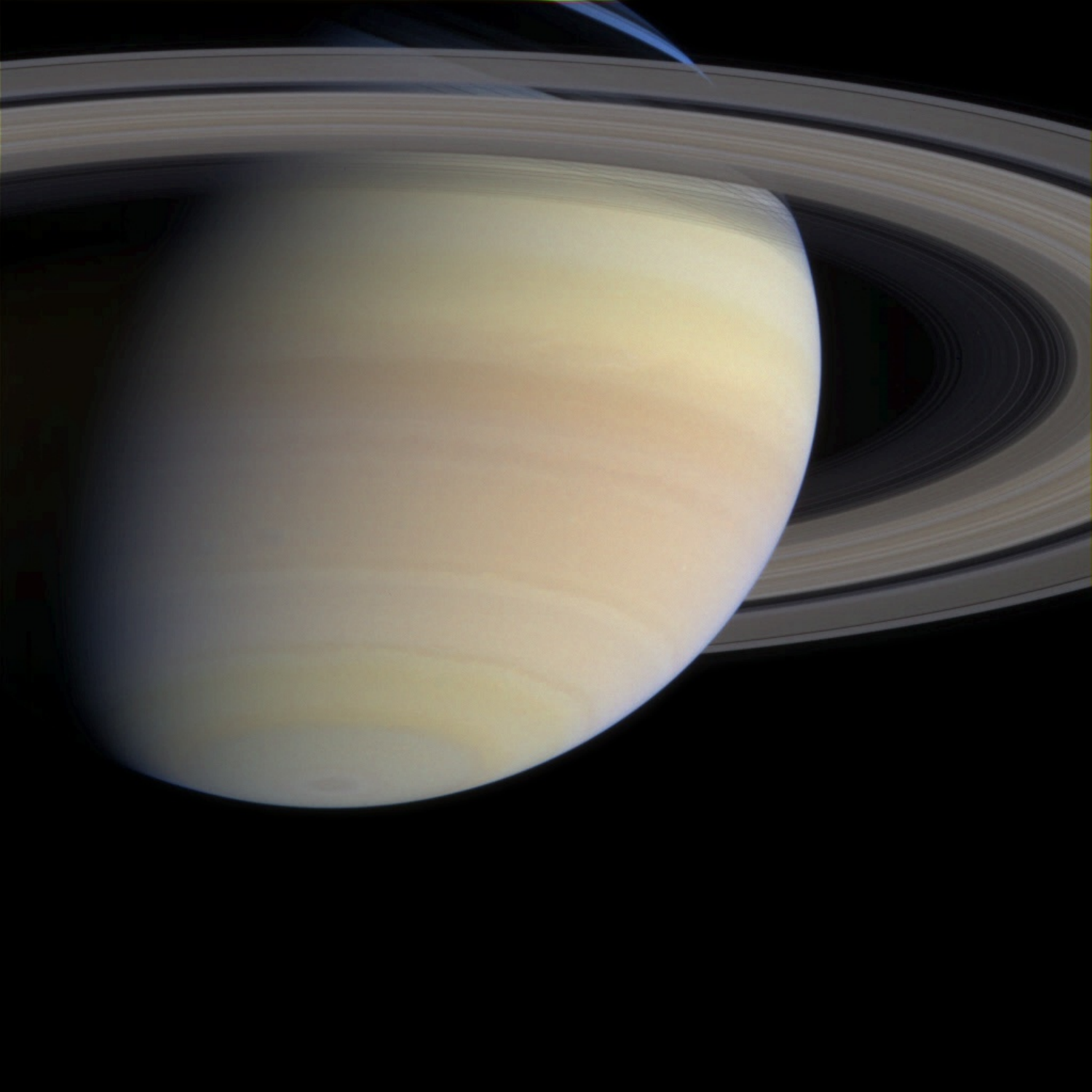}
\caption{This true-color image was obtained in 2004 by the Cassini ISS camera on approach to Saturn.  It illustrates the strong hemispheric color difference observed during southern summer.  Fig. \ref{saturn_montage} shows how Saturn's colors evolved with time during the Cassini mission. Courtesy NASA/JPL-Caltech.}
\label{sat2004}
\end{figure}

The seasonal asymmetry in tropospheric aerosols was therefore well established prior to Cassini's arrival just after southern summer solstice.  The tropospheric haze optical thickness was expected to be the largest and most extended in the summer hemisphere, and smallest in the winter hemisphere, with the transition occurring at some time near to the equinox as previously observed by Voyager ($L_s=8-18^\circ$) and Pioneer ($L_s=354^\circ$).  The Cassini spacecraft arrived at Saturn at an earlier seasonal phase ($L_s\approx290^\circ$, southern summer), and has now provided the opportunity to track these changing hemispheric asymmetries.  Indeed, Cassini observations pre-equinox revealed a  hemispheric asymmetry that was opposite to those seen by Voyager post-equinox.  As shown in Fig. \ref{sat2004}, the high northern latitudes showed a vibrant blue color in 2004.  The interpretation \citep{12edgington} is that Rayleigh scattering by gas molecules is responsible, and that the colored haze material is suppressed in the northern (winter) latitudes relative to southern (summer) latitudes.  Subsequent Cassini ISS images have shown that the blue color persisted into 2008 but by 2009 (near equinox) the blue color had dissipated at northern latitudes (see the images in Fig. \ref{saturn_montage}).  At the current epoch (2014) the southern high latitudes are beginning to show a blue color as they recede into winter conditions.  These observations are consistent with the idea that the blue color indicates reduced production of haze throughout the winter season.

\begin{figure}%
\figurebox{3.2in}{}{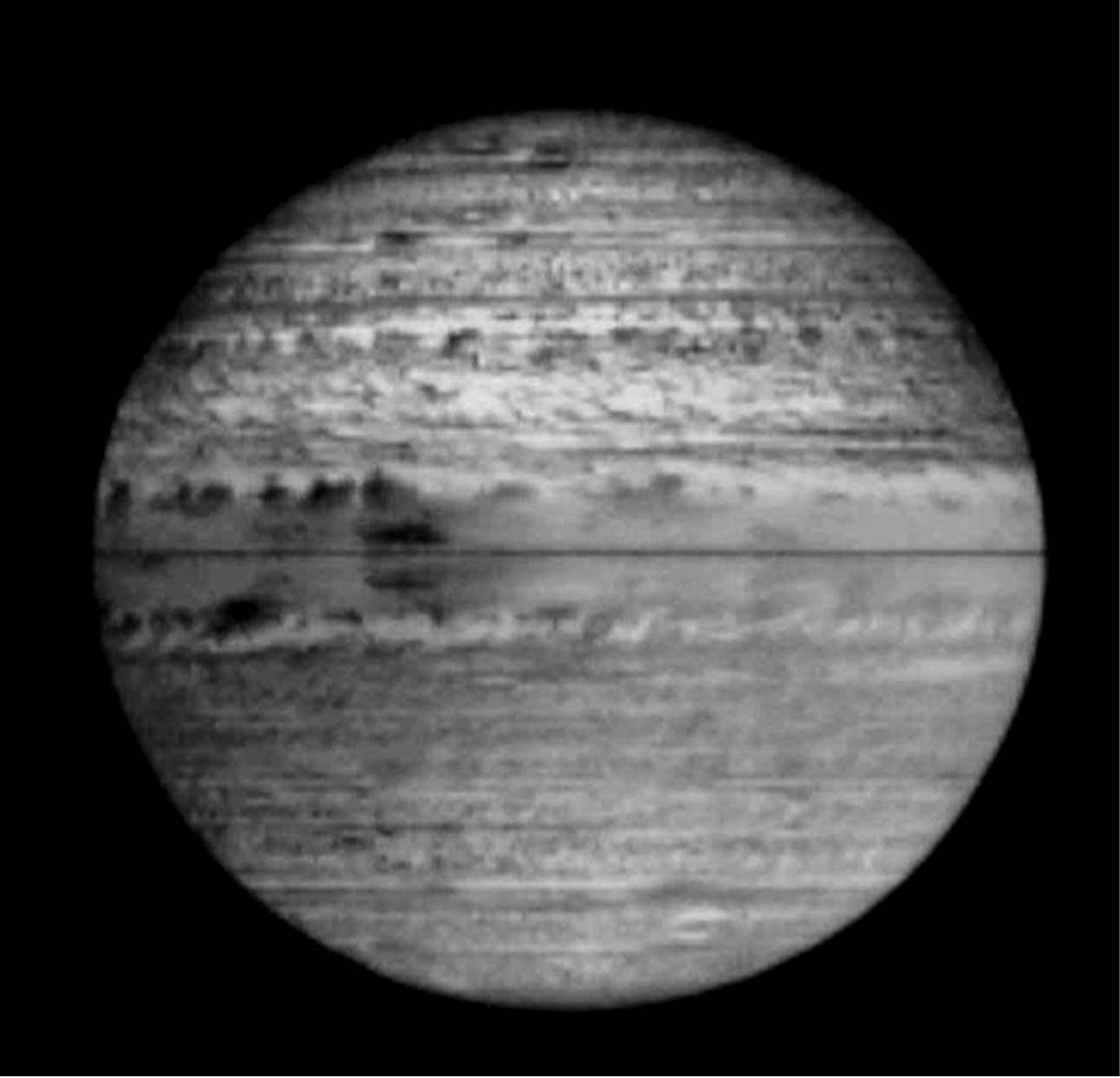}
\caption{Saturn at 5 microns wavelength, showing cloud features in silhouette against the bright background of Saturn's own thermal emission \citep{06baines_dps, 09west}.   In addition to fine-scale banding and numerous dynamic features, note the hemispheric differences in brightness and contrast, caused by a thicker tropospheric haze in the southern hemisphere. Courtesy NASA/JPL-Caltech.}
\label{satVIMS}
\end{figure}

\begin{figure}%
\figurebox{3.2in}{}{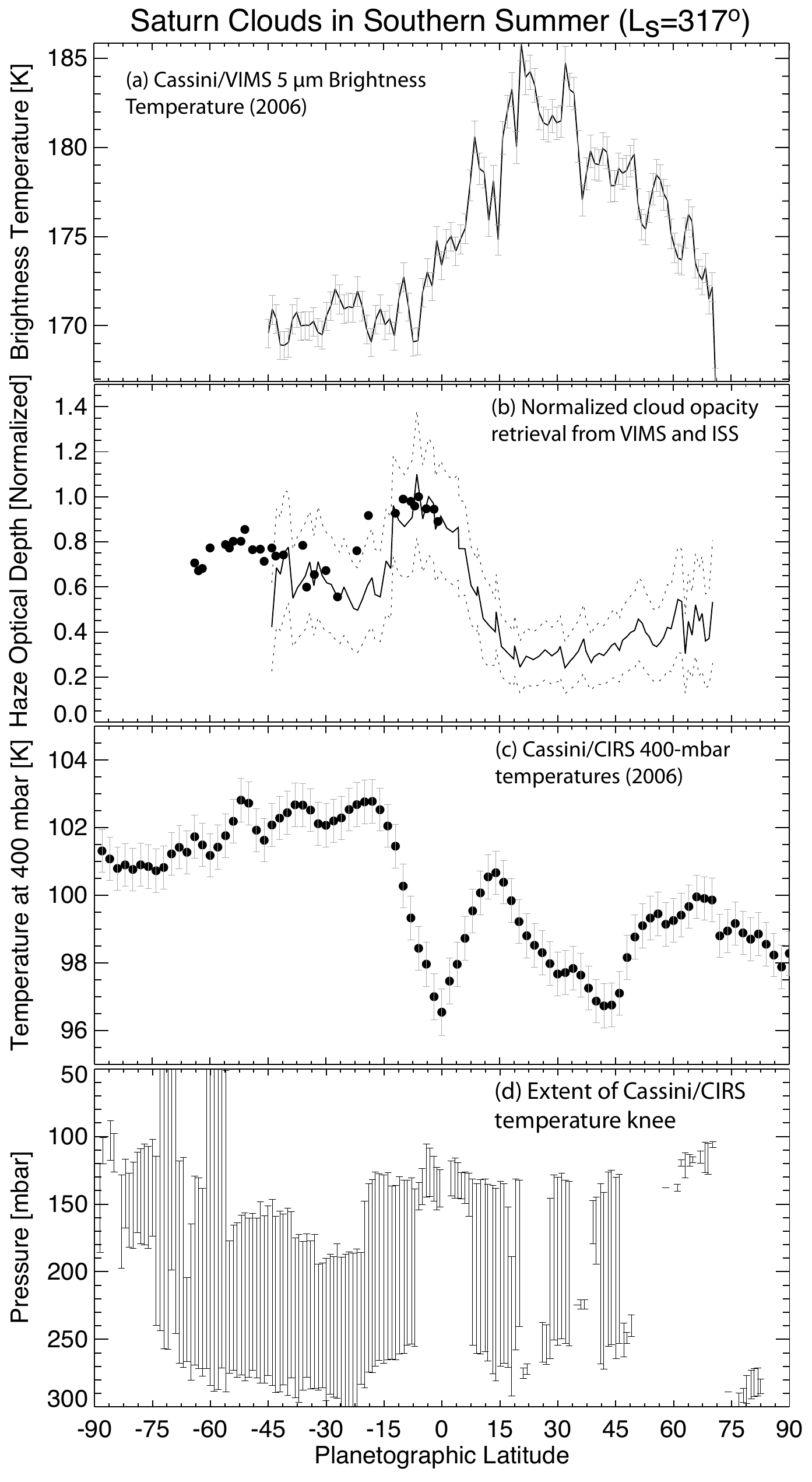}
\caption{Comparison of the observed 5-$\mu$m asymmetry in brightness temperature in southern summer in panel a (2006, $L_s=317^\circ$) with upper tropospheric haze opacities derived from Cassini/ISS for the southern hemisphere \citep[black circles,][]{13roman} and Cassini/VIMS for both hemispheres \citep[solid line with dotted error range,][]{11fletcher_vims}.  These are found to be largely correlated with the upper tropospheric thermal structure (400 mbar temperatures from Cassini/CIRS) in panel c.  Furthermore, the asymmetry correlates with the extent of the temperature `knee' observed by CIRS in the upper troposphere (panel d), and believed to be caused by localized heating in the tropospheric haze layer \citep{07fletcher_temp}. }
\label{cloud_compare}
\end{figure}

These inferences from the Cassini/ISS visible data are consistent with infrared observations from Cassini/VIMS, particularly at 5 $\mu$m where the dearth of hydrogen and methane opacity permits the escape of radiation from relatively deep (3-6 bar) levels, and cloud opacity serves to attenuate this 5-$\mu$m flux.  As first reported by \citet{06baines_dps}, Fig. \ref{satVIMS} shows that Saturn's northern hemisphere was brighter than the southern summer hemisphere, with southern hemisphere contrasts muted due to the relatively higher aerosol opacity overlying the contrast-producing clouds.  \citet{11fletcher_vims} performed a quantitative analysis of Cassini/VIMS cubes from April 2006 ($L_s=317^\circ$), finding opacity in two regimes:  a compact cloud deck centered in the 2.5-2.8 bar region, symmetric between the two hemisphere with small-scale opacity variations responsible for the numerous light/dark axisymmetric lanes; and secondly a hemispherically asymmetric population of aerosols at $p<1.4$ bar which was $\approx2.0\times$ more opaque in the southern summer hemisphere.  The upper tropospheric haze asymmetry is shown in Fig. \ref{cloud_compare}b, compared to the CIRS-derived 400-mbar temperatures (panel c) and the observed contrast in 5-$\mu$m brightness temperature (panel a).  The vertical structure of this upper-level `haze' could not be constrained by the nightside VIMS observations, but is likely to be the same material responsible for the homogenous haze observed in CCD methane-band imaging.  The deep cloud was at higher pressures than the predicted condensation altitude for NH$_3$ \citep[1.8 bar for a $5\times$ enrichment of heavy elements,][]{99atreya}, but at lower pressures than the predicted levels for NH$_4$SH condensation (5.7 bar), so its composition could not be identified unambiguously.  Unfortunately, there are no studies of the expected reversing of the 5-$\mu$m cloud asymmetry available in the literature today.

\citet{13roman} conducted a quantitative study of Cassini/ISS images of the southern summer hemisphere between 2004-2007 ($L_s=296-333^\circ$), reproducing the data with a stratospheric haze merging into a tropospheric haze that sits within the convectively-stable region of the upper troposphere (i.e., above the R-C boundary discussed in Section \ref{temp}).  The tropospheric haze was found to reach the greatest heights ($40\pm20$ mbar) at the equator, but to sit deeper ($140\pm20$ mbar) at southern mid-latitudes.  Fig. \ref{cloud_compare} compares the southern hemisphere optical depths per bar at 619 nm derived by \citet{13roman} with the haze opacities at 5 $\mu$m derived from \citet{11fletcher_vims}.  We have normalized to the maximum haze opacity at the equator to allow intercomparison, highlighting the north-south asymmetry and the comparison with the CIRS-derived temperature structure.  The haze location derived by \citet{13roman} is consistent with the localized aerosol heating observed in Saturn's thermal field by \citet{07fletcher_temp} (Fig. \ref{cloud_compare}d), and \citet{13roman} suggest that the tropospheric haze is correlated with Saturn's mid-summer temperatures (Fig. \ref{cloud_compare}c).  At higher pressures, the authors find discrete cloud structures in the 1-2 bar range, which may or may not be the same as the 2.5-2.8 bar cloud inferred from VIMS.  Quantitative ISS studies have yet to extend into the northern winter hemisphere, and resolving the apparent discrepancy in cloud vertical structure with VIMS is a source of ongoing activity.

Finally, \citet{13sromovsky} analyzed VIMS reflectivity observations in the vicinity of the northern storm, and found that the main haze (of particles 1 $\mu$m in radius or smaller) in Saturn's northern hemisphere in 2011 (away from the storm-perturbed regions) was located between 111-178 mbar (top) and 577-844 mbar (bottom), depending on the latitude, with a deep, compact and opaque cloud near 2.6-3.2 bar.  This is broadly consistent with the 5-$\mu$m thermal emission studies.  They confirmed that this haze contained no spectroscopically-identifiable features of pure condensates in the VIMS spectral range, and no signatures of hydrazine (N$_2$H$_4$), although diphosphine cannot be definitively ruled out.  But, as for ISS, the latitudinal and seasonal dependence of Saturn's reflectivity has not yet been investigated.

In summary, the historical record of Saturn's aerosol distributions has shown that seasonal insolation changes induce hemispheric asymmetries in the tropospheric (and potentially stratospheric) hazes, with higher opacity in the summer hemisphere and low opacity (and blue color) in the winter hemisphere (Fig. \ref{cloud_compare}).  The haze sits in the region approximately between the tropopause and the radiative-convective boundary, above the main convective region.  Cassini images are showing that the asymmetry is reversing, along with the upper tropospheric temperatures, although quantitative studies of VIMS and ISS reflectivity have only been published for single epochs.  

\section{Conclusions and outstanding questions}
\label{conclude}

The longevity and broad wavelength coverage of the Cassini mission, coupled with the decades-long record of ground-based observations, have revealed intricate connections between Saturn's atmospheric temperatures, chemistry and aerosol formation mechanisms.   Environmental conditions in the stably-stratified upper troposphere (approximately situated above the radiative-convective boundary at 400-500 mbar) and stratosphere have been observed to vary over time in response to the shifting levels of solar energy deposition and the efficiency of radiative cooling to space.  Atmospheric temperatures have tracked the seasonal insolation changes, albeit with a phase lag that increases with depth into the atmosphere; the upper tropospheric haze has changed in optical thickness, causing differences in the coloration of reflected sunlight in the summer and winter hemispheres; and the zonal mean distributions of both tropospheric and stratospheric gaseous constituents exhibit hemispheric asymmetries that may be subtly shifting with time.  The atmospheric soup of gases and aerosols in turn affects the radiative properties of the atmosphere (i.e., the rates of heating and cooling), which further influences the seasonal temperature shifts that we observe.  Atmospheric circulation and localized dynamics can redistribute energy and material from place to place, implying that thermal and chemical perturbations are superimposed onto the large-scale seasonal asymmetries and sometimes (in the case of equatorial uplift and vertical waves; or within the polar vortices) can dominate the observed spatiotemporal trends.

Disentangling all of these competing effects is the key challenge for the next generation of modeling activities, towards a complete simulation of the seasonal behavior of Saturn's cloud-forming weather layer, upper troposphere and stratosphere.  Historically, models have been developed in isolation to explain one subset of the larger problem - for example, radiative climate simulations (with or without convective adjustment and advection of heat via circulation) have been used to understand the magnitude of the seasonal temperature changes; one-dimensional photochemical modeling with parameterized vertical mixing (and an absence of horizontal mixing and circulation) demonstrate how stratospheric hydrocarbons vary with time and location; and equilibrium cloud condensation models predict where key condensates should be forming in Saturn's troposphere.  However, each of these models should be intricately linked, as gaseous and aerosol distributions influence the temperatures (and vice versa), which influences chemical and cloud microphysics time scales, as well as dynamic redistribution of heat.  The latest generation of numerical simulations are moving in this direction - for example, \citet{12friedson} combined radiative modeling with atmospheric circulation; \citet{15hue} connects the photochemically-predicted hydrocarbon distributions with seasonal temperature changes in the context of the model of \citet{10greathouse}; and \citet{14spiga_dps} is aiming to incorporate the radiative model of \citet{14guerlet} into a full general circulation model.  However, some key pieces of the puzzle (the radiative effect of poorly understood and seasonally-variable aerosols; the influence of atmospheric circulation) continue to elude the community and remain the subject of ongoing theoretical development.  

To date, Cassini has monitored Saturn's complex atmosphere for only a third of a Saturnian year.  The upper tropospheric and stratospheric temperature fields have been measured from late northern winter through to late northern spring, which happens to overlap with Voyager observations just after the previous northern spring equinox, three decades ago.  Surprisingly, the atmospheric temperatures measured by Cassini and Voyager \textit{at the same point in the seasonal cycle} are not identical \citep{13li, 14sinclair, 15fletcher_fp1}, suggesting that Saturn might experience non-seasonal variability in its thermal field and circulation.  Another mystery is why the stratospheric seasonal response at very low pressures ($p<0.1$ mbar) appears to be more muted than that at 1 mbar, counter to the expectations of radiative climate models in the absence of stratospheric circulation.  Atmospheric temperatures will continue to be monitored in the thermal-infrared using Cassini until northern summer solstice, and afterwards with ground-based and space-based observations (e.g., JWST), albeit restricted to the earth-facing summer hemisphere.  Long-term consistent datasets are essential to confirm whether Saturn truly does undergo non-seasonal variability.  

Compared to the study of Saturn's temperature field, measurements of the distribution of gaseous and aerosol species are less mature.  Cassini has identified hemispheric asymmetries in tropospheric species (para-H$_2$, the disequilibrium species PH$_3$ and the condensible volatile NH$_3$, Section \ref{trop_chem}), tropospheric haze opacities and cloud coloration (Section \ref{clouds}) and photochemically-produced stratospheric hydrocarbon species (ethane, acetylene, propane, diacetylene and methylacetylene, see Section \ref{strat_chem}).  It remains to be seen whether the timescales for the processes generating these asymmetries match or exceed a Saturnian year.  If the timescales are short, then we may expect to see some reversal of the asymmetries as northern summer solstice approaches.  If the timescales are long (potentially related to the southern-summer timing of perihelion and the northern-summer timing of aphelion) then we might expect the asymmetries to be quasi-permanent features of Saturn's atmosphere.  Temporal studies of these distributions have only been published for the two principle hydrocarbons (ethane and acetylene) and the tropospheric para-H$_2$ fraction, showing slow and subtle changes to their distributions that may be more influenced by atmospheric circulation than by seasonally-variable production and loss (e.g., a seasonally-reversing Hadley cell at the equator; inter-hemispheric transport from the autumn to the spring hemisphere; and strong subsidence over the north polar region that has recently emerged into spring sunlight).  In the near future, new data from Cassini will hopefully determine the magnitude of seasonal shifts in the tropospheric and stratospheric haze distributions, in addition to the higher-order hydrocarbons and the tropospheric species.  By completing Cassini's observational characterisation of Saturn's seasonal atmosphere through to northern summer solstice, we hope to inform and guide the development of the next generation of numerical simulations to establish Saturn as the paradigm for seasonal change on a giant planet.

\section*{Acknowledgements}  

The authors wish to thank A. Laraia, J. Hurley, J. Sinclair, J. Friedson and M. Roman for sharing the results of their studies.  The manuscript benefited from thorough reviews by K. Baines, R. Achterberg, F.M. Flasar and G. Bjoraker.  Fletcher was supported by a Royal Society Research Fellowship at the University of Oxford and University of Leicester.

\bibliography{references}\label{refs}
\bibliographystyle{cambridgeauthordate-Leigh}

\backmatter



 \printindex

\end{document}